\documentclass[twocolumn]{aastex701}

\usepackage{amsmath}
\newcommand{\EDIT}[1]{\textcolor{black}{#1}}

\begin{document}

\title{A New Scaling Law for Non-Dipolar Magnetic Fields in Rapidly Rotating Stars and Planets}

\author[orcid=0000-0002-6788-8898,sname='Wulff']{Paula N. Wulff}
\affiliation{Department of Earth, Planetary, and Space Sciences,\\ 
University of California, Los Angeles (UCLA), 595 Charles E Young Dr East, Los Angeles, CA 90095, USA}
\email[show]{paulawulff@epss.ucla.edu} 

\author[orcid=0000-0002-6917-8363,sname='Cao']{Hao Cao}
\affiliation{Department of Earth, Planetary, and Space Sciences,\\ 
University of California, Los Angeles (UCLA), 595 Charles E Young Dr East, Los Angeles, CA 90095, USA}
\email{hcao@epss.ucla.edu} 

\author[orcid=0000-0002-8642-2962,sname='Aurnou']{Jonathan M. Aurnou}
\affiliation{Department of Earth, Planetary, and Space Sciences,\\ 
University of California, Los Angeles (UCLA), 595 Charles E Young Dr East, Los Angeles, CA 90095, USA}
\email{jona@epss.ucla.edu} 

\begin{abstract}

Magnetic field generation in giant planets and rapidly rotating stars produces a diverse range of field geometries, from large-scale dipole-dominated configurations to complex, small-scale multipolar structures. Earlier dynamo studies have suggested that multipolar solutions tend to arise when rotational effects become less dominant. We investigate the strength of non-dipolar magnetic fields generated in systems dominated by rotation. 40 three-dimensional, spherical-shell dynamo simulations were carried out using the MagIC code, primarily made up of bistable pairs - simulations with the same control parameters that can settle in both a dipolar and non-dipolar steady-state regime. We use this suite of models to test how their magnetic field strength scales with heat flux and velocity. Our dynamo simulations produce magnetic fields with morphologies that fall on the two distinct branches, dipolar or non-dipolar, yet have very similar convective velocities. The strength of the dipole component differs by an order of magnitude between the two regimes, when scaled as a function of driving power. However, their non-dipolar magnetic field strengths are very similar. We conclude that when attempting to predict the magnetic field strength of rapidly rotating planets and stars, one cannot assume that it will have a dipole-dominated geometry. In particular, the amplitude of the dipole component is expected to be an order of magnitude smaller in the non-dipolar regime.

\end{abstract}


\section{Introduction}\label{sec:Intro}

Planetary and stellar magnetic fields are generated by dynamo action in which convective motions of electrically conducting fluids sustain magnetic fields against ohmic dissipation \citep{Moffatt_2978, Roberts_2000, Moffatt_2019}. Identifying robust scaling laws that relate the amplitude and morphology of these fields to underlying control parameters, such as rotation rate and heat flux, has been a central goal of dynamo theory. In rapidly rotating systems—relevant to Earth, the gas giants, and many low-mass stars—the Coriolis force strongly constrains fluid motions, often leading to large-scale magnetic fields with a dominant axial dipole component \citep[e.g.][]{Christensen_2006, Aubert_2009, Calkins_2021}. Scaling arguments, based on a balance between the ohmic dissipation associated with the magnetic field and the energy flux available to drive dynamo action, have successfully described aspects of dipole-dominated dynamos, providing benchmarks that connect numerical models with planetary and stellar observations \citep[e.g.][]{Christensen_2006, Christensen_2009, Christensen_2010, Davidson_2013}.

However, many natural dynamos operate outside this dipolar regime. Non-dipolar, or multipolar, fields are common, ranging from the complex morphologies of Uranus and Neptune \citep{Ness_1986, Connerney_1991, Holme_1996} to the variable large-scale fields observed in rapidly rotating M dwarfs \citep[e.g.][]{Morin_2010, Shulyak_2017, Bellotti_2024}. Numerical studies have revealed that the transition from dipolar to multipolar regimes is controlled by a delicate balance between buoyancy forcing, rotation, and magnetic feedbacks \citep{Menu_2020, Tassin_2021, Zaire_2022, Soderlund_2025}. Yet the scaling behaviour of non-dipolar dynamos remains less well constrained. Very few previous studies consider whether the laws derived in dipolar contexts extend to multipolar fields, or whether distinct force balances and scaling regimes emerge \citep{Yadav_2013, Yadav_2013b}.

In this study, we investigate dynamo scaling laws in the rapidly rotating regime with a particular focus on non-dipolar field configurations. Using numerical dynamo simulations across a broad range of Rayleigh and Ekman numbers, we quantify how magnetic energy, morphology, and characteristic length scales depend on buoyancy flux and rotation rate. We focus on the energy flux scaling laws of \citet{Christensen_2006} and \citet{Christensen_2009} (we will later refer to the non-dimensional form of this law as [CA06], and the dimensional form as applied to stars as [C09]). By systematically comparing dipolar and non-dipolar cases, we evaluate the universality of the existing scaling law and identify necessary extensions to account for multipolar magnetic fields.

Unlike previous studies, we perform separate analyses of the dipole component in both regimes. \EDIT{First, the dipole is the most distinguishing difference between the magnetic fields in the two regimes. It is further motivated by the fact that it is the component that can be most reliably determined by observations}. For all Solar System planets, and for most exoplanets, the dynamo region does not extend to the surface. As a result, even for planetary magnetic fields measured \textit{in-situ}, we can accurately determine only their large-scale components. This is particularly true for Uranus and Neptune, given the single fly-by Voyager II measurements. The small-scale components lose power rapidly before reaching the planetary surface. Consequently, it becomes difficult to test any scaling relation for the total magnetic field strength without making assumptions about how it relates to the observed large-scale field. Moreover, many key phenomena linked to a star’s or planet’s magnetic field outside the body—such as the stellar wind or a planetary magnetosphere—depend primarily on the dipole component. This further underscores the importance of developing reliable methods for estimating it.

Our results aim to provide a more comprehensive framework for connecting dynamo theory with the observed diversity of planetary and stellar magnetic fields.

\section{Methods} \label{sec:Methods}

Systems where rotation plays a significant role often feature dipolar magnetic fields. Analysing non-dipole dominated magnetic field generation in the same regimes requires exploring a parameter space where there is a bistability of the two states. Thus, we base our modelling approach on one of the previous studies where multiple bistable cases were identified \citep{Yadav_2013} and expand the dual (dipolar-multipolar) parameter space from there. Numerous bistable dynamo states have also been observed in studies using the anelastic approximation \citep[e.g.][]{Gastine_2012} and including an anelastic radial density contrast is no doubt closer to reality for most planets and stars. However, as this is the first study to explicitly focus on multipolar magnetic fields in rotation-dominated dynamo models, we choose a Boussinesq approach to lay down the groundwork before introducing more complexity.

\subsection{Governing Equations}

Our numerical set-up consists of a spherical shell which rotates along the $z$-axis and which is bounded by inner radius $r_i$ and outer radius $r_o$. The aspect ratio $r_i/r_o$ is 0.35. A linear variation of gravity with radius is assumed. We non-dimensionalize the magnetohydrodynamic (MHD) equations by using the shell thickness $r_o - r_i=D$ as the reference length scale and viscous diffusion time $\tau_\nu=D^2/\nu$, where $\nu$ is the fluid viscosity, as the time unit. The magnetic field B is scaled by $\sqrt{\rho\mu\lambda\Omega}$, where $\rho$ is the constant fluid density, $\mu$ is the magnetic permeability, $\lambda$ the constant fluid magnetic diffusivity, and $\Omega$ is the rotation rate. The velocity is scaled by the Reynolds number $Re=uD/\nu$.

The temporal evolution of velocity $\boldsymbol{u}$, temperature $T$, and magnetic field $\boldsymbol{B}$ is described by the magnetohydrodynamic (MHD) equations under the Boussinesq approximation:
\begin{align}
    \frac{\partial \boldsymbol{u}}{\partial t}& + \boldsymbol{u} \cdot\nabla \boldsymbol{u} + \frac{2}{E} \hat{e}_z \times \boldsymbol{u} + \nabla P \nonumber\\
    &=\frac{Ra}{Pr}g(r)T \hat{e}_r+ \frac{1}{EPm}(\nabla\times B) \times B + \nabla^2\boldsymbol{u},\\
    \frac{\partial T}{\partial t} &+ \boldsymbol{u} \cdot\nabla T =\frac{1}{Pr} \nabla^2T, \\
    \frac{\partial \boldsymbol{B}}{\partial t}&= \nabla\times (\boldsymbol{u} \times \boldsymbol{B}) +\frac{1}{Pm}\nabla^2\boldsymbol{B},\\
    \nabla\cdot \boldsymbol{u} &= 0, \\
    \nabla\cdot  \boldsymbol{B} &= 0.
\end{align}
This system of equations is governed by several non-dimensional control-parameters: Ekman number $E = \nu/\Omega D^2$; the Rayleigh number $Ra =\alpha g_o D^3\Delta T/\nu\kappa$, where $g_o$ is gravity at the outer boundary, $\alpha$ is the thermal expansivity, and $\kappa$ is the thermal conductivity; magnetic Prandtl number $Pm=\nu/\lambda$; and Prandtl number $Pr =\nu/\kappa$. The control parameter values used in our suite of simulations are tabulated in the Appendix.

We also consider the convective Rossby number $Ro_c=\sqrt{RaE^2/Pr}$, which estimates the ratio of buoyancy and Coriolis forces \citep[e.g.][]{Aurnou_2020}. When $Ro_c\ll1$, we consider a system to be in the rapidly rotating regime. The $Ro_c>1$ regime is not explored in this study.

We assume free-slip mechanical boundaries at both inner and outer radius. The magnetic field matches a potential field at both boundaries. A fixed temperature contrast $\Delta T$ is maintained between the top and the bottom. 

\subsection{Numerical Methods} \label{sec:NumMethod}

We use the open-source three-dimensional MHD code MagIC \citep[\url{https://magic-sph.github.io/},][]{Christensen_2001, Wicht_2002, Gastine_2012b} where Chebyshev polynomials are used in the radial direction and spherical harmonic decomposition in the horizontal directions.

The first models at $E=10^{-4}$ and $Ra=6\times10^6$, $Ra=8\times10^6$ and $Ra=1\times10^7$ are started from random temperature fluctuations and either i) a strong dipolar magnetic field \EDIT{\citep[as is common practice for geodynamo studies, e.g.][]{Frasson_2026}} or ii) a small-scale magnetic field, which will lead the solution to land in the dipolar or multipolar regimes respectively \EDIT{(see Appendix~\ref{sec:Init} for more details)}. Higher Rayleigh and lower Ekman number simulations are initiated from this first set of models, to reduce convergence time and save computing resources.

The time-averaging for the diagnostic parameters described in Section \ref{sec:Diag} is started after the initial transient has passed and a steady state is reached. It is performed for a minimum of $0.1\,\tau_\nu$.
   
\subsection{Diagnostic Parameters}\label{sec:Diag}

\begin{table}[h!]
\caption {Definitions of the various forms of the Lorentz number used in this work. $S_s$ is the spherical shell surface and $\mathcal{EM}(r_o)$ is the surface integrated total magnetic field strength (at the outer boundary). We also include the definition of the volume- and surface-averaged dipolarities, $f_{dip,V}$ and $f_{dip,r_o}$, respectively. Here, $\mathcal{P}_B(\ell=1)$ being the power in the magnetic dipole. \label{tab:LoDefs}} 
\centering
\begin{tabular}{ll}
\hline\hline             
Parameter & Definition\\
\hline\hline   
$Lo$ & $B/(\sqrt{ \mu_0\rho}\Omega D)$ \\
\hline
$Lo_o$ & $E(2\langle \mathcal{EM}(r_o)\rangle_t\,/S_s)^{1/2}$ \\
$Lo_o(\ell=1)$ & $E(2\langle f_{dip,r_o} \,\mathcal{EM}(r_o)\rangle_t\,/S_s)^{1/2}$ \\
$Lo_o(\ell>1)$ & $E(2\langle (1-f_{dip,r_o}) \,\mathcal{EM}(r_o)\rangle_t\,/S_s)^{1/2}$ \\
\hline
$Lo_V$ & $E(2\langle \mathcal{EM}\rangle_t\,/V_s)^{1/2}$ \\
$Lo_V(\ell=1)$ & $E(2\langle f_{dip,V} \,\mathcal{EM}\rangle_t\,/V_s)^{1/2}$ \\
$Lo_V(\ell>1)$ & $E(2\langle (1-f_{dip,V})\, \mathcal{EM}\rangle_t\,/V_s)^{1/2}$ \\
\hline
$f_{dip,V}$ & $\mathcal{P}_B(\ell=1)\,/\sum^{\ell_{max}}_{\ell=1}\mathcal{P}_B(\ell)$ \\
$f_{dip,r_o}$ & $\mathcal{P}_B(\ell=1, r_o)\,/\sum^{\ell_{max}}_{\ell=1}\mathcal{P}_B(\ell, r_o)$
\end{tabular}
\end{table}

To assess whether a model is in the rapidly rotating regime, we use the previously defined convective Rossby number $Ro_c$ as well as the \textit{a posteriori} local Rossby number $Ro_\ell$, which is a diagnostic to estimate the importance of inertia relative to rotation in a dynamical system.

The kinetic energy $\mathcal{EK}$ in each system is used to obtain flow amplitudes as a function of time. Here we consider the convective components, from both the poloidal and toroidal non-axisymmetric parts of $\mathcal{EK}$, to obtain the system-scale Rossby number $Ro$ ($=u^\prime/\Omega D$):
\begin{align}
    \mathcal{EK}_{conv} &=\frac{1}{2}\int\boldsymbol{u}^\prime\cdot\boldsymbol{u}^\prime \,dV,\\
    Ro &=E\left(2\frac{\mathcal{EK}_{conv}}{V_s}\right)^{1/2},\label{eq:Rop}
\end{align}
where $V_s$ is the volume of the shell and primes denote the non-axisymmetric components of the velocity. We also evaluate the characteristic non-dimensional convective flow length-scale:
\begin{equation}
    \frac{\mathcal{L}_U}{D}=\pi\left(\sum_\ell \ell\frac{\boldsymbol{u}^\prime_\ell\cdot\boldsymbol{u}^\prime_\ell}{\boldsymbol{u}^\prime\cdot\boldsymbol{u}^\prime}\right)^{-1},
\end{equation}
where $\ell$ is the spherical harmonic degree and $\boldsymbol{u}^\prime_\ell$ is the non-axisymmetric velocity field at the corresponding degree. We can then obtain the time-averaged local Rossby number of each model:
\begin{equation}\label{eq:Rol}
    Ro_\ell=\langle Ro D/\mathcal{L}_U\rangle_t,
\end{equation}
where $\langle...\rangle_t$ denotes averaging over time. The magnetic energy (including both the poloidal and toroidal components) is used to calculate the total magnetic field strength in each system:
\begin{equation}
    \mathcal{EM} = \frac{1}{2}\int\boldsymbol{B}\cdot\boldsymbol{B}\,dV.
\end{equation}
Following \citet{Christensen_2006, Yadav_2013} we express the non-dimensional magnetic field strength in terms of the Lorentz number
\begin{equation}\label{eq:Lo}
    Lo=B/(\sqrt{ \mu_0\rho}\Omega D),
\end{equation}
which is evaluated for our models using the expressions given in Table~\ref{tab:LoDefs}. This can be related to the often-used Elsasser number, expressing the ratio of Lorentz to Coriolis forces, through $\Lambda=Lo^2 Pm/E$. While the volumetric Lorentz number $Lo_V$ is the classical parameter used in previous works, to make our analysis more comparable with measurements we also evaluate surface Lorentz number values $Lo_o$, also defined in Table~\ref{tab:LoDefs}. 

The table also gives the definitions for dipolarity, where we again define the surface, $f_{dip,r_o}$, and the volumetric, $f_{dip,V}$, values. In general, $f_{dip,r_o}$ is the quantity we refer to when classifying a model as dipolar or non-dipolar. Similarly to previous studies, we typically find that this value is fairly constant for dipolar models while it fluctuates much more for non-dipolar systems. This is information we lose when performing a time-average, yet for the purposes of this study it allows for a simple, classical, characterisation.

The [CA06] scaling law, relating magnetic field strength to available driving power, is expressed in terms of the non-dimensional parameters $Lo$ and the flux-based Rayleigh number $Ra^*_Q$ \citep{Christensen_2006, Yadav_2013}:
\begin{equation}\label{eq:CA06}
    Lo/\sqrt{f_{ohm}}\propto{Ra^*_Q}^{1/3},
\end{equation}
where $f_{ohm}$ is the ratio of ohmic dissipation to total dissipation. The flux-based Rayleigh number, 
\begin{equation}
    Ra^*_Q = \frac{1}{4\pi r_or_i}\frac{\alpha g_o Q_{adv}}{\rho c_P \Omega^3D^2},
\end{equation}
(where $c_P$ is the heat capacity) is used rather than the $\Delta T$ based Rayleigh number, as this can be estimated more easily for stars and planets by using approximations of their advected heat flux $Q_{adv}$.

In terms of our model diagnostics this is equivalent to:
\begin{equation}
    Ra^*_Q = \frac{(Nu-1)Ra E^3}{Pr^2},
\end{equation}
where $Nu$ is the Nusselt number.

However, we can also frame the scalings in terms of Rossby numbers, using the asymptotic scaling relations for rapidly (or slowly) rotating systems \citep{Julien_2012, Aurnou_2020}. In terms of the convective Rossby number we have
\begin{align}
    \frac{(Nu-1)}{(Ra Pr)^{1/2}} \sim \begin{cases}
            Ro_c^2, \text{ for } Ro_c \ll 1\\
            \text{const.}, \text{ for } Ro_c \gg 1.
           \end{cases}
\end{align}
Further, asymptotic estimates of the convective flow velocities (see eq.~\ref{eq:Rop}) yield:
\begin{align}
    Ro \sim \begin{cases}
            Ro_c^2, \text{ for } Ro_c \ll 1\\
            Ro_c, \text{ for } Ro_c \gg 1,
           \end{cases}
\end{align}
for the system-scale Rossby number, and the local Rossby number (see eq.~\ref{eq:Rol}):
\begin{equation}
    Ro_\ell \sim Ro_c.
\end{equation}
These scalings lead to
\begin{align}
    Ra^*_Q \sim \begin{cases}
            Ro_c^5\sim Ro^{5/2}\sim Ro_\ell^5, \text{ for } Ro_c \ll 1 \label{eq:Rocscal}\\
            Ro_c^3\sim Ro^3\sim Ro_\ell^3, \text{ for } Ro_c \gg 1.
           \end{cases}
\end{align}
Using the convective Rossby number for our numerical framework has the advantage of $Ro_c$ being an input parameter for simulations, as well as being able to easily judge whether one is in an inertia ($Ro_c \gg 1$) or rotation ($Ro_c \ll 1$) dominated regime. However, $Ro_c$ is difficult to estimate for planets and stars as it requires approximating the convective velocities, or convective turnover times. Thus, we present the scaling laws in terms of $Ra^*_Q$ in the results section, but include scalings with various Rossby numbers in Appendix~\ref{sec:LoRoScale}.
   

   \begin{figure}[h!]
   \section{Results}
   \centering
   \includegraphics[width=\hsize]{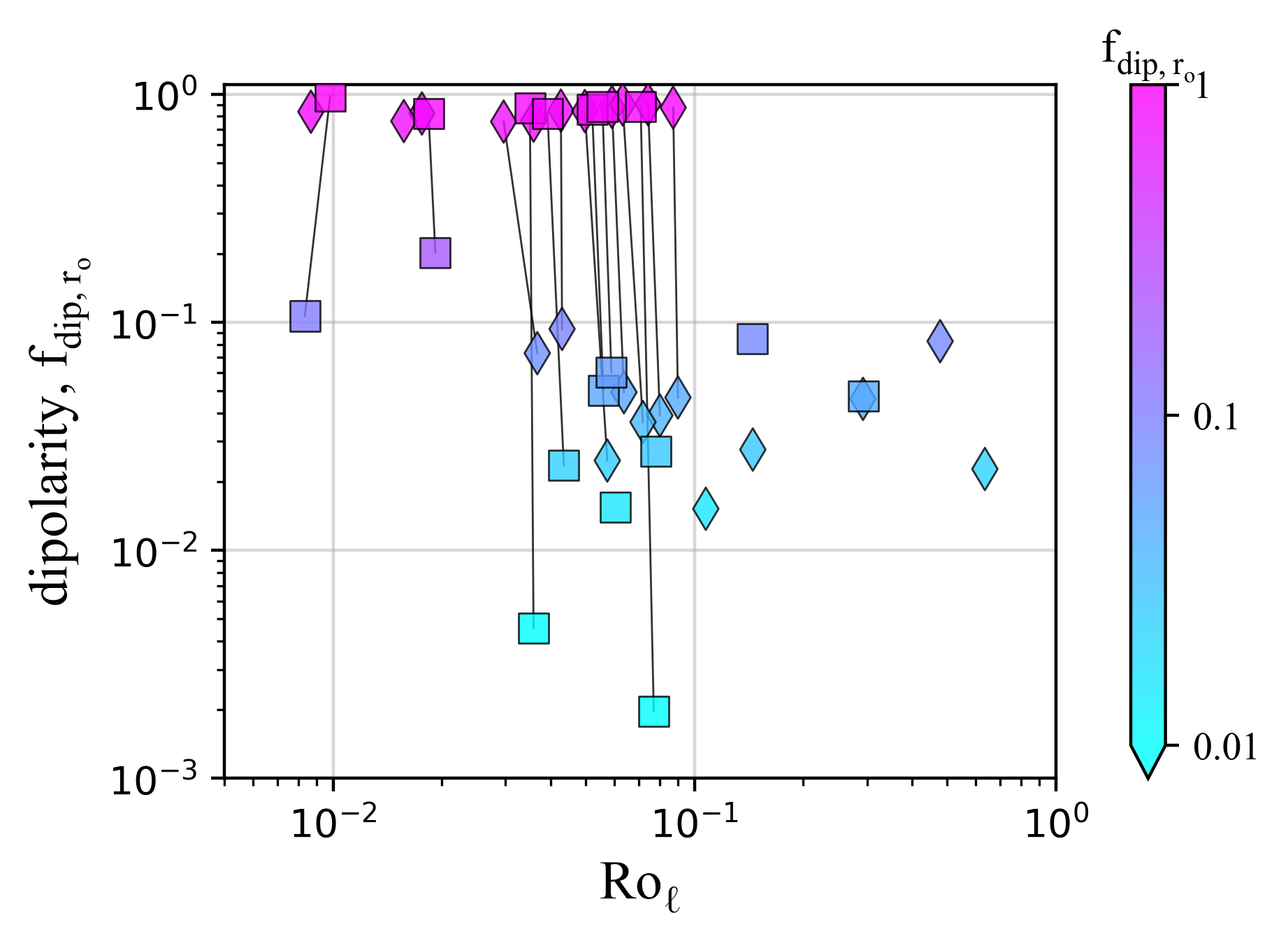}
      \caption{Surface dipolarity as a function of local Rossby number. Bistable pairs (dipolar and non-dipolar systems which have identical control parameters) are connected by black lines. Diamonds (squares) represent models with $Pm=1$ ($Pm=0.5$).\label{fig:fdipRol}}
   \end{figure}

      \begin{figure*}[t!]
   \centering
   \includegraphics[width=\hsize]{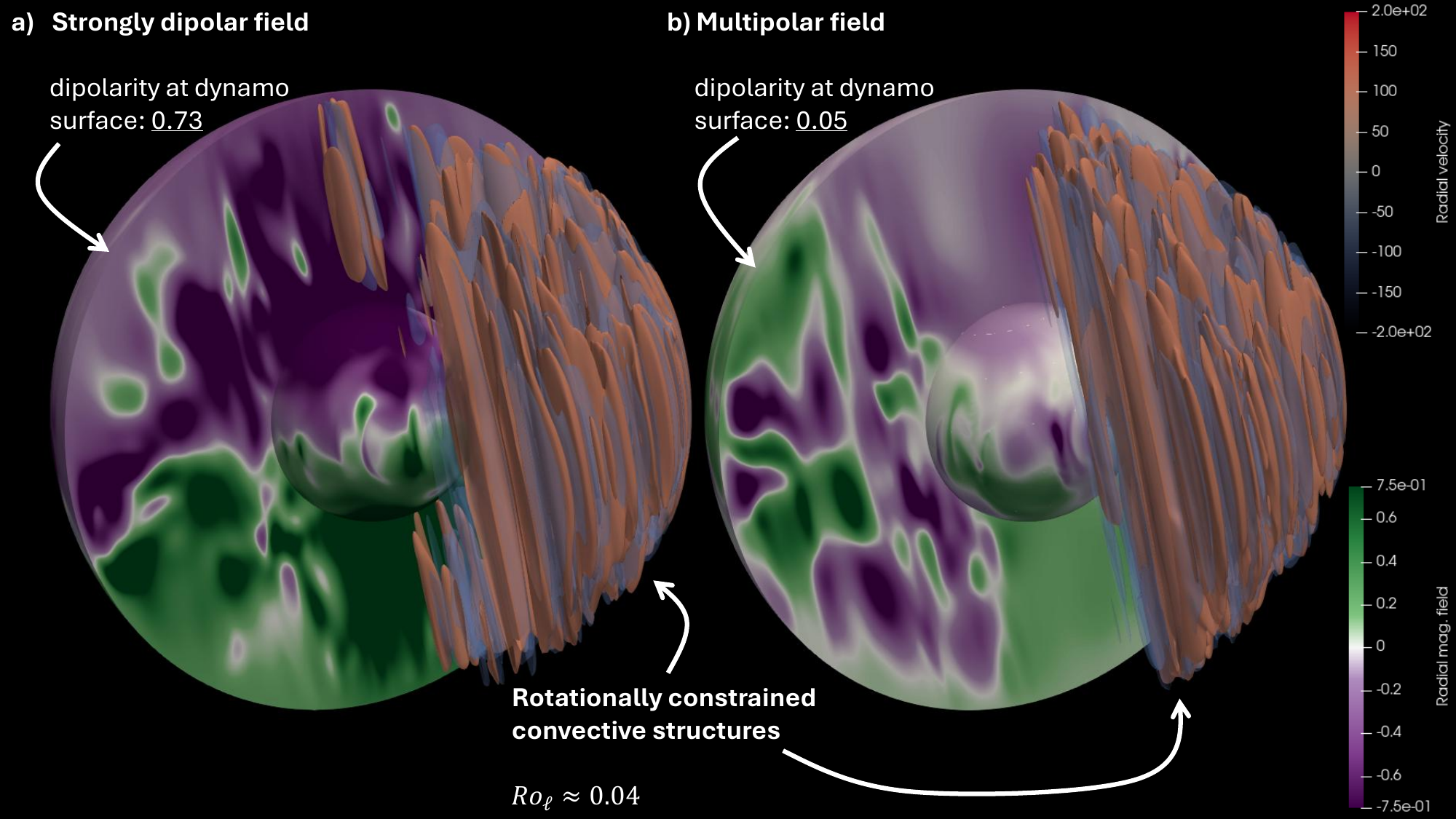}
      \caption{Three-dimensional visualisations of a) dipolar and b) non-dipolar model\EDIT{s}, both with $E=1\times10^{-5}$, $Ra=2\times10^8$, and $Pm=0.5$. Isocontours of outwards/inwards radial flows are shown in red/blue. Radial magnetic field is shown in green/purple.}
         \label{fig:3Dviz}
   \includegraphics[width=\hsize]{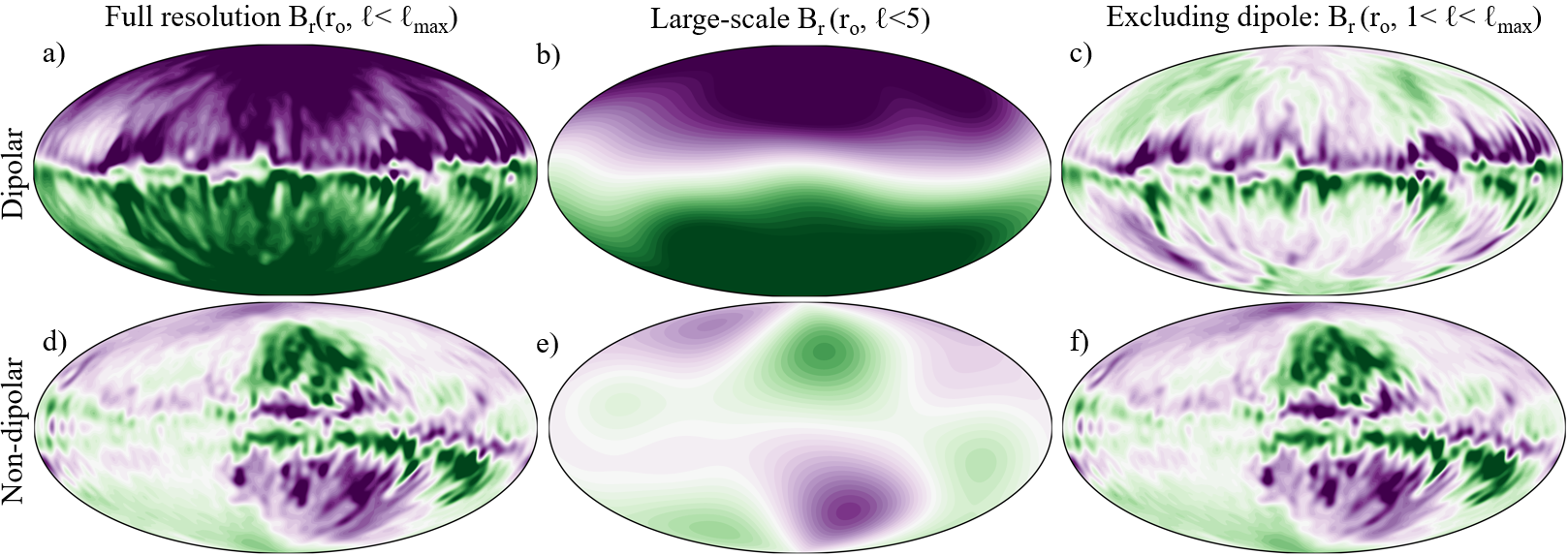}
      \caption{Surface radial magnetic fields for a a-c) dipolar and d-f) non-dipolar model, both with $E=1\times10^{-5}$, $Ra=2\times10^8$, and $Pm=0.5$. a) and d) use the full resolution, b) and e) include only degrees $\ell<5$ (mirroring our current resolution of the ice giants' magnetic fields), and c) and f) exclude the dipole component. The same colour-scale of $\pm0.25$ (in Elsasser units) is used in all panels.}
         \label{fig:2Dviz}
   \end{figure*}
   
Fig.~\ref{fig:fdipRol} shows the surface dipolarity as a function of local Rossby number (Equation \ref{eq:Rol}) for all simulations in this study. The bistable regime is seen to extend from the lowest local Rossby numbers we reach to around $Ro_\ell=0.1$. 
The lines connecting bistable pairs in Fig.~\ref{fig:fdipRol} highlight that these models not only have the same input parameters and therefore the same value of $Ro_c$, but the \textit{diagnostic} local Rossby number is also very similar.
   
   \begin{figure}[h!]
   \centering
   \includegraphics[width=0.95\hsize]{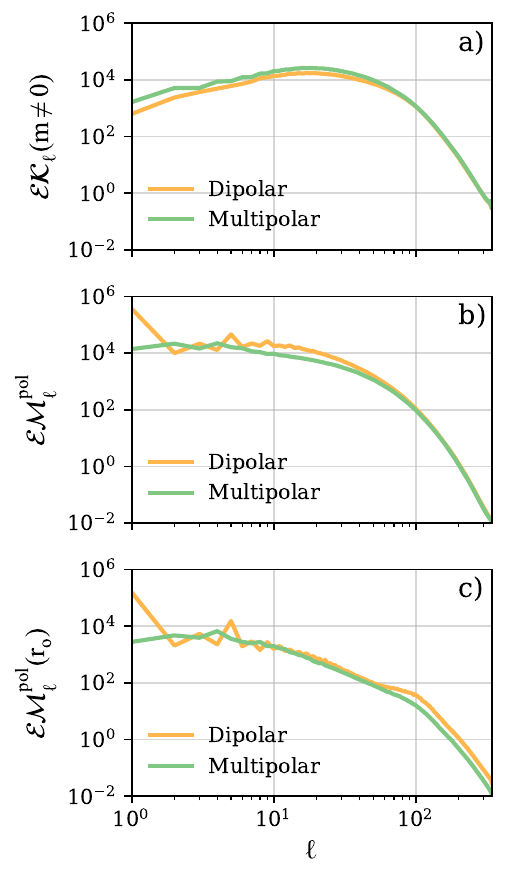}
      \caption{Energy spectra of a bistable pair of models ($E=10^{-5}$, $Ra=2\times10^8$, $Pm=0.5$, $Ra/Ra_c=19.8$, $Ro_c=0.14$). $\mathcal{EK}(m\ne0)$ refers to the poloidal and toroidal non-axisymmetric components of the time-averaged kinetic energy spectra. $\mathcal{EM}^{pol}$ ($\mathcal{EM}^{pol}(r_o)$) refers to the time-averaged, volumetric (surface) poloidal magnetic energy spectra. \EDIT{Both models have $\ell_{max}=341$.} \label{fig:Spectra}}
   \end{figure}

A three-dimensional visualization of bistable models ($E=10^{-5}$, $Ra=2\times10^8$, $Pm=0.5$, $Ro_c=0.14$) is shown in Fig.\ref{fig:3Dviz}a and b. This highlights the similarity between the convective flow structures, illustrated by the red/blue isocontours, which are strongly aligned with the axis of rotation ($Ro_\ell\approx0.04$ for this pair). In contrast, the magnetic field structures differ dramatically, as the dipole component dominates over the smaller scales for the dipolar model. This can be seen more clearly by comparing Fig.~\ref{fig:2Dviz}a and d, which show snapshots of the fully resolved radial magnetic fields at the surface of the two models. The two differ dramatically. Fig.~\ref{fig:3Dviz}c and f show the same field snapshots but excluding the $\ell=1$ component. They highlight how insignificant the dipole contribution is for the non-dipolar system.

The spatial distribution of the kinetic and magnetic energy of one pair of the same pair of models is shown in terms of their spectra in Fig.~\ref{fig:Spectra}. We plot a) the total non-axisymmetric kinetic energy (representative of the convection) and both the b) volumetric and c) surface spectra of the poloidal component of the magnetic energies, as a function of spherical harmonic degree $\ell$. Fig.~\ref{fig:Spectra}a shows that the convection is slightly stronger in the multipolar case for this particular bistable pair of models, indicating that the stronger magnetic field of the dipolar model is suppressing the convection more than in the multipolar system. Apart from this small difference in amplitude (up to $\ell\sim70$) the shape of the spectra are very similar. We observe that the shape and amplitude of the magnetic energy spectra are very similar for both models - with the stark exception of the dipole, $\ell=1$, component. The only other noticeable differences are in the volumetric poloidal magnetic energy between $7\lesssim\ell\lesssim40$ and in the surface energy for $\ell\gtrsim80$, where there is more energy at those scales in the dipolar model.


\subsection{Rossby Number Scaling}

   \begin{figure}[t!]
   \centering
   \includegraphics[width=0.95\hsize]{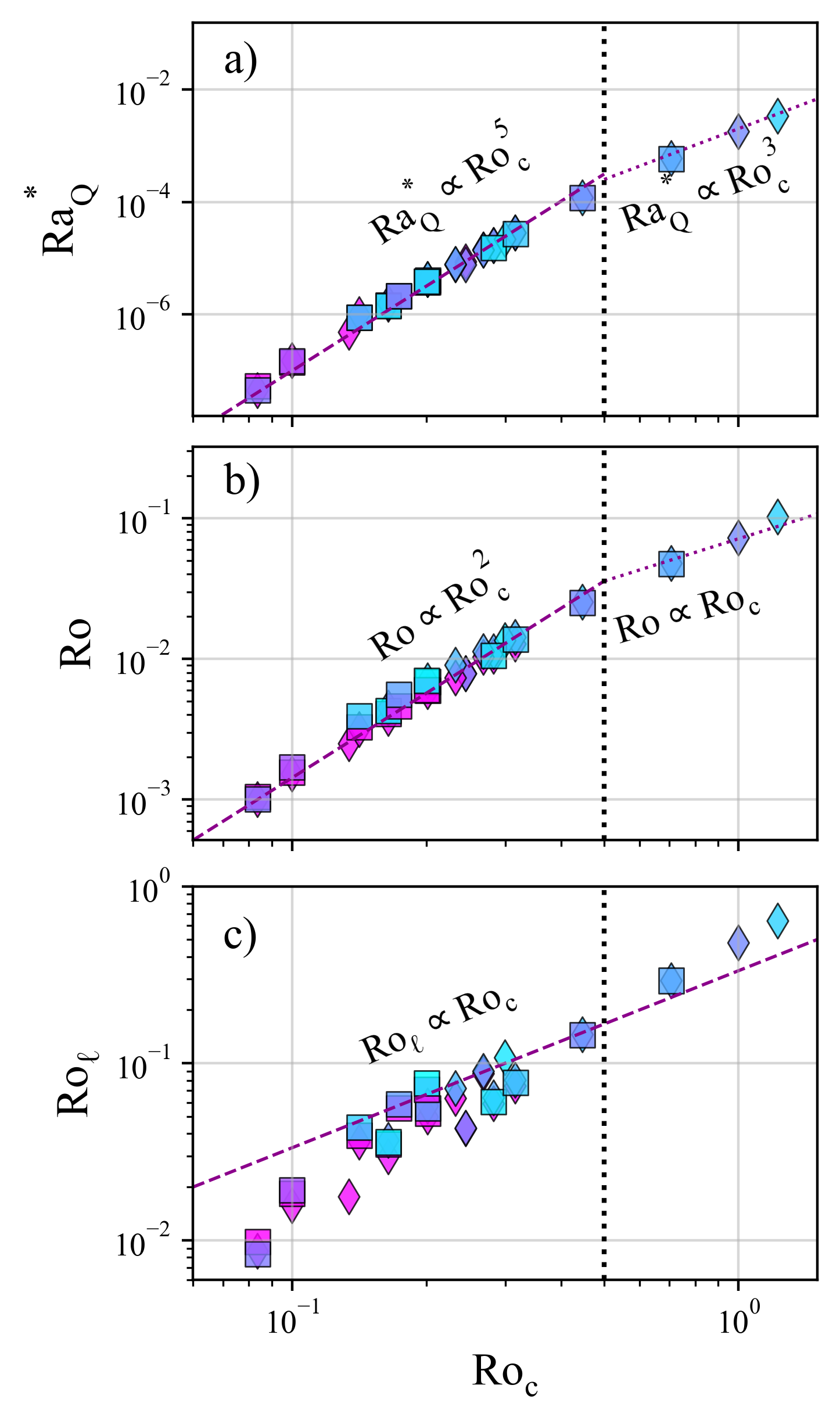}
      \caption{$Ra^*_Q$, $Ro$ and $Ro_\ell$ as a function of convective Rossby number, with the expected rapidly/slowly rotating scalings indicated by dashed/dotted purple lines (see Eq.~\ref{eq:Rocscal}). Symbol colours represent dipolarity, as in Fig.~\ref{fig:fdipRol}. The vertical dotted black line denotes the limit of the data we use to make our fittings, $Ro_c<0.5$, as we observe clear deviation from the rapidly rotating limit beyond this point.}
         \label{fig:Roc_RaQ_Rop_Rol}
   \end{figure}
   
As described in Sec.~\ref{sec:Diag} and summarised in equation~\ref{eq:Rocscal}, we have predictions of how the three forms of the Rossby number introduced here ($Ro_c$, $Ro$ and $Ro_\ell$) should scale with respect to each other, and the flux-based Rayleigh number $Ra^*_Q$. We show how well these scalings apply to our set of models in Fig.~\ref{fig:Roc_RaQ_Rop_Rol}.

While we do not plot $Ro(Ra^*_Q)$ directly, Fig.~\ref{fig:Roc_RaQ_Rop_Rol} shows that both $Ra^*_Q$ and $Ro$ follow the expected scaling with $Ro_c$ very well. Only the $Ro_\ell\sim Ro_c$ scaling shows a little more shingling of the data points, with more super-critical models better hitting the trend line. This could perhaps be improved if a different definition of the length-scale $\mathcal{L_U}$ were used, but we did not explore this further here.

\cite{Christensen_2006} make a fit of $Ro\propto {Ra^*_Q}^{0.41}$ (although this is calculated using the total kinetic energy, i.e. including zonal and meridional flows). Meanwhile, \cite{Yadav_2013, Yadav_2013b} make separate fittings for the non-axisymmetric $Ro$, obtaining exponents of 0.44 and 0.42, respectively. Thus, all three studies agree well with the asymptotically predicted value of $2/5$ (see Eq.~\ref{eq:Rocscal}). 


\begin{table}[h!]
\subsection{Magnetic Field Strength Scaling}
\caption {Coefficients, $c$ and their associated $1\sigma$ values when fitting equation~\ref{eq:CA06} for various forms of $Lo$ for the dipolar (D), multipolar (M) and universal (U) regimes. The third column gives the coefficient and $Ra^*_Q$ exponent when making no assumption of the power law. We exclude simulations with $Ro_c>0.5$ from the fitting.\label{tab:RaQFits}} 
\centering
\begin{tabular}{lll}
\hline\hline             
$Lo$, regime & ${Ra^*_Q}^{1/3}$ coeff. & Free power fitting \\
\hline\hline   
$Lo_o$, D & 0.168 $\pm$ 0.022 & $(0.60 \pm 0.12)\,{Ra^*_Q}^{0.44 \pm 0.01}$ \\
$Lo_o$, M & 0.075 $\pm$ 0.008 &$(0.32 \pm 0.11)\,{Ra^*_Q}^{0.47 \pm 0.03}$ \\
$Lo_o(\ell=1)$, D & 0.157 $\pm$ 0.022 & $(0.58 \pm 0.11)\,{Ra^*_Q}^{0.45 \pm 0.01}$\\
$Lo_o(\ell=1)$, M & 0.015 $\pm$ 0.003 & $(0.03 \pm 0.03)\,{Ra^*_Q}^{0.41 \pm 0.07}$\\
$Lo_o(\ell>1)$, U & 0.069 $\pm$ 0.006 & $(0.30 \pm 0.08)\,{Ra^*_Q}^{0.47 \pm 0.02}$\\
\hline
$Lo_V$, D & 0.646 $\pm$ 0.072 & $(1.26 \pm 0.41)\,{Ra^*_Q}^{0.39 \pm 0.02}$\\
$Lo_V$, M & 0.471 $\pm$ 0.061 & $(0.94 \pm 0.26)\,{Ra^*_Q}^{0.40 \pm 0.02}$\\
$Lo_V(\ell=1)$, D & 0.559 $\pm$ 0.067 & $(0.96 \pm 0.28)\,{Ra^*_Q}^{0.38 \pm 0.02}$\\
$Lo_V(\ell=1)$, M & 0.069 $\pm$ 0.019 & $(0.04 \pm 0.04)\,{Ra^*_Q}^{0.30 \pm 0.08}$\\
$Lo_V(\ell>1)$, U & 0.424 $\pm$ 0.052 & $(1.64 \pm 0.45)\,{Ra^*_Q}^{0.46 \pm 0.02}$\\
\hline
\end{tabular}
\end{table}

   \begin{figure*}[t!]
   \centering
   \includegraphics[width=\hsize]{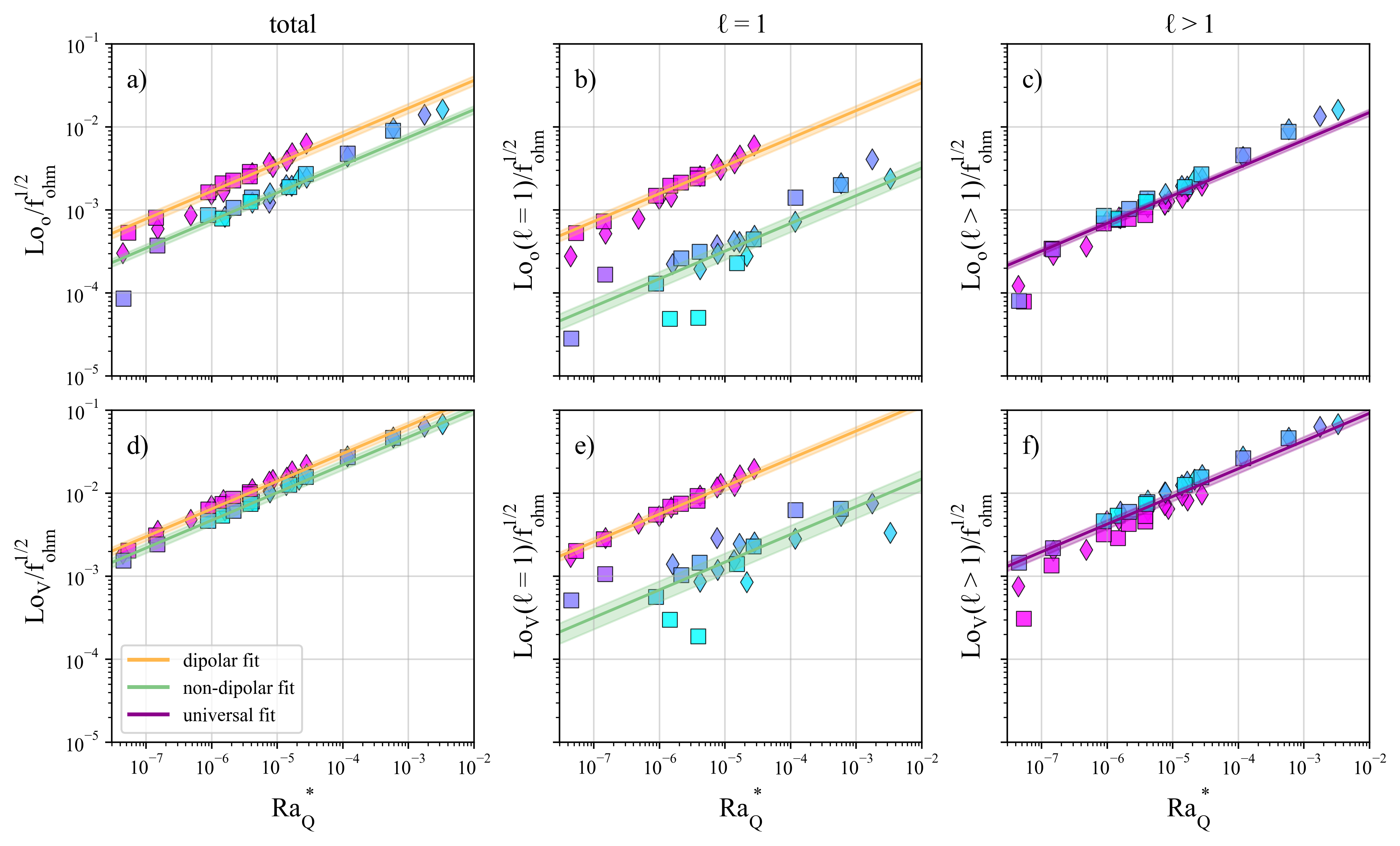}
\caption{Lorentz number, \EDIT{adjusted by the ohmic dissipation fraction,} as a function of the modified Rayleigh number based on the advected heat flux. Panels a)-c) show surface values of $Lo$ while panels d)-f) show volumetric values. a) and d) show total $Lo$, b) and e) $Lo(\ell=1)$, and c) and f) $Lo(\ell>1)$ (see Table~\ref{tab:LoDefs}). Diamonds (squares) represent models with $Pm=1$ ($Pm=0.5$), while colour indicates surface dipolarity: dipolar/non-dipolar in magenta/cyan (see colour-bar of Fig.~\ref{fig:fdipRol}). Orange/green lines in a), b), d) and e) show dipolar/non-dipolar fits to $Lo/f_{ohm}^{1/2}\sim{Ra^*_Q}^{1/3}$ while purple lines in d) and f) show seem Table~\ref{tab:RaQFits}).
\label{fig:LoRaQ}}
\end{figure*}

\EDIT{We fit our data to equation~\ref{eq:CA06} and the value of the coefficient} thus obtained is given for each of the variations of $Lo$ described in sec.~\ref{sec:Diag} in Table~\ref{tab:RaQFits}. We give the fitting when assuming the exponent is $1/3$, the exponent which emerges from theory \citep[e.g.][]{Davidson_2013} and also from dimensional analysis, as well as the scaling when the exponent is left as a fitting parameter. In general, we find that the fitted exponent is a little greater than $1/3$, particularly when evaluated from the surface Lorentz number values. The multipolar fitting for $Lo_V(\ell=1)$ marks the exception to this, and has an exponent of only $0.30\pm0.08$. The steeper slope for the surface $Lo$ fitting could be attributed to the weak magnetic field strengths of our lowest $Ra_Q^*$ cases, which sit slightly below the rest (see Fig.~\ref{fig:LoRaQ}a and b). This may be due to these systems being fairly close to the onset of convection. 

Comparing the panels of Fig.~\ref{fig:LoRaQ}, we find that for the total, volume averaged, magnetic field (panel d) the exponential pre-factors are fairly similar ($0.646\pm0.072$ and $0.471\pm0.061$ for dipolar/non-dipolar models, respectively) yet the two sets are distinct. This is similar to what was found in \citet{Yadav_2013, Yadav_2013b}, although \EDIT{their pre-factors of $0.9$ and $0.7$, respectively, differ a little from ours.} 
Fig.~\ref{fig:LoRaQ}a shows that this difference between dipolar and non-dipolar models is enhanced at the surface. When focusing exclusively on the dipole component of the models (panels b and e), we find an even more significant difference between the regimes (which is to be expected) and the pre-factors differ by an order of magnitude, both in the volume and at the surface. We also observe slightly more scatter for the non-dipolar models around the line of best-fit, which reflects the fact that the dipole component does not play a significant role in the dynamics of these systems and can therefore vary more (also with time, as mentioned in Sec.~\ref{sec:Diag}).

When comparing the strength of the non-dipole components of the two regimes, we find that they follow the same scaling (shown in Fig.~\ref{fig:LoRaQ}c and f), in particular at the surface, $Lo_o(\ell>1)$. 
This is consistent with what the spectra in Fig.~\ref{fig:Spectra} indicated. This finding suggests that the energy in all $\ell>1$ magnetic field components is directly set by the available driving power, irrespective of whether the regime is dipolar or non-dipolar.

\section{Magnetic Field Estimates}

We now explore how our newly obtained scaling laws may be applied to objects inside and outside our solar system.

\subsection{M Dwarfs}

   \begin{figure*}[t]
   \centering
   \includegraphics[width=\hsize]{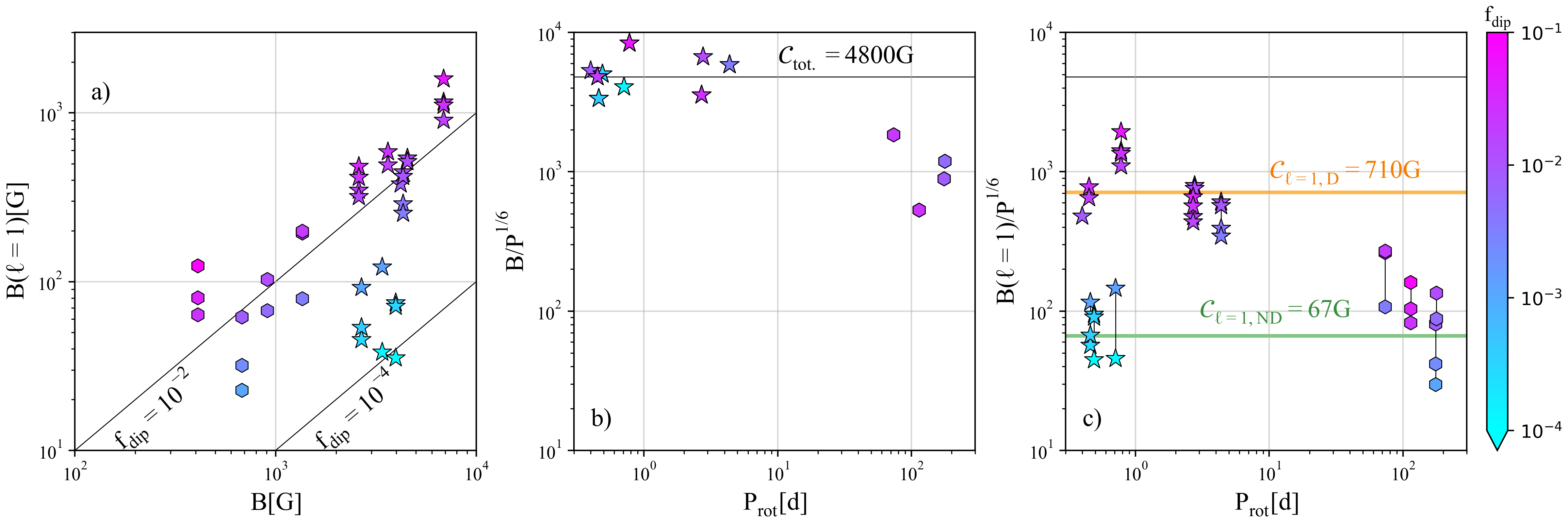}
      \caption{a) Average (poloidal) dipole field strength $B(\ell=1)$ as a function of average total field strength $B$. b) $B$, and c) $B(\ell=1)$, relative to the power $P\sim ML^2/R^7$ according to the [C09] scaling law (Equation \ref{eq:C09}), as a function of rotation period. All stars in our sampled are fully convective ($M_\star<0.35M_\odot$). \EDIT{See Appendix~\ref{sec:MDwarfData}, Table~\ref{tab:MDwarfs}, for data sources.} 
      As the same stars have multiple measurements for $B(\ell=1)$ we join all markers for the same star with a black line. $M$, $L$ and $R$ to calculate $P$, as well as $P_{rot}$, are taken from \citet{See_2025} - see this paper for original sources. Symbol colour indicates dipolarity and star markers represent rapidly rotating bodies while hexagons represent those with $P_{rot}>10$d. On b) we indicate the coefficient of $\mathcal{C}_{tot}=4,800\,$G while on c) we indicate the average ratio $\mathcal{C}_{\ell=1}=B(\ell=1)/P^{1/6}$, for dipolar and non-dipolar separately, excluding the slow rotators.}
         \label{fig:MDwarfs}
   \end{figure*}
   
We harvest magnetic field data for M dwarfs from several studies, obtained using different observational methods.

For the large scale magnetic field components we consult the composite study of \citet{See_2025}. We focus on fully convective stars, with $M<0.35M_\odot$ as we can expect stars with a significant radiative region to deviate from our scaling laws that are based on fully convective spherical shell models. This leads us to restrict our data harvesting to \citet{Morin_2008a, Morin_2008b, Morin_2010, Bellotti_2024, Lehmann_2024}\footnote{We exclude the 2019 $B(\ell=1)$ estimate for EV Lac from \citet{Bellotti_2024} as they comment that the magnetic field strength for this epoch is likely underestimated.}, where magnetic field strengths are obtained via Zeeman-Doppler Imaging, ZDI \citep[as well as principal component analysis in][]{Bellotti_2024, Lehmann_2024}, relying on circularly polarised light - Stokes V signatures. This is necessary because only the Stokes V signatures are sensitive to magnetic field topology.

From these studies we take the average (poloidal) dipole field strength from ZDI, $\langle B_{dip}\rangle$, which we name $B(\ell=1)$ in our framework. 

However, small-scale magnetic field features with opposite polarities are cancelled out when using the Stokes V, so it does not capture a large fraction \citep[$\sim90\%$,][]{Reiners_2009} of the total magnetic energy. Therefore, for the total magnetic field strength we consider \citet{Shulyak_2017, Reiners_2022}, who analyse unpolarised light - Stokes I signatures. These give us the global average magnetic field strength $B$.
   
These two measurements are shown in Fig.~\ref{fig:MDwarfs}a (and tabulated, including the respective star names in Appendix~\ref{sec:MDwarfData}). To quantify dipolarity as consistently as possible with our numerical study, we estimate
\begin{equation}
    f_{dip}\approx B(\ell=1)^2/B^2.
\end{equation}
This is indicated by the marker-colour in Fig.~\ref{fig:MDwarfs}. However, we use different limits for our colour-scale compared to our numerical analysis. This is due to stars having a much larger fraction of their magnetic energy in smaller scales compared to our simulations. We attribute this to not being able to reach high enough magnetic Reynolds numbers, $Rm$, with the limitations we have on computing power - for stars $Rm\sim\mathcal{O}(10^8)$ while for gas planets $Rm\sim\mathcal{O}(10^4-10^5)$. Thus, for these objects we consider $f_{dip}=3\times10^{-3}$ as our classification boundary. 

This is also why, when considering the total average magnetic field strength, we only see a vague indication that for non-dipolar magnetic fields $B$ lies slightly under the predicted value \citep[as was noticed by][]{Shulyak_2017} - see Fig.~\ref{fig:MDwarfs}b. The dipole component makes up much less of the total field, even in the dipolar systems, than in our models and therefore barely impacts the total average magnetic field strength. This is therefore very similar for both regimes (see fig.~\ref{fig:LoRaQ}c and f).

To test how these measurements fit with respect to our power-based scaling laws:
\begin{equation}\label{eq:C09}
    B^2\propto P^{1/3},
\end{equation}
we estimate the driving power available as derived by \citet{Christensen_2009, Reiners_2009b},
\begin{equation}
    P= \left(\frac{M}{M_\odot}\right)\left(\frac{L}{L_\odot}\right)^2\left(\frac{R}{R_\odot}\right)^{-7},
\end{equation}
where we omit a pre-factor, for now. Here, $M$ is stellar mass, $L$ is its bolometric luminosity and $R$ is its radius. In Fig.~\ref{fig:MDwarfs}b and c we plot $B$ and $B(\ell=1)$, respectively, divided by $P^{1/6}$ as a function of rotation period $P_{rot}$.

Fig.~\ref{fig:MDwarfs}b shows that the total magnetic field strength of the rapidly rotating stars in our sample is estimated very well by the [C09] scaling law with
\begin{equation}
    B_{pred}=4,800 \text{ G }\times P^{1/6}.\label{eq:MDwarfTot}
\end{equation}
For stars with $P_{rot}<10\,$d there is no dependence on rotation rate, as the scaling law predicts.

In fig.~\ref{fig:MDwarfs}c we show the same stars but now study their average poloidal dipole magnetic field strength. We see that rapidly rotating stars with dipolar ($f_{dip}>3\times10^{-3}$) and non-dipolar ($f_{dip}<3\times10^{-3}$) magnetic field topologies cluster around two fairly distinct regions in $B(\ell=1)/P^{1/6}$ space. Taking an average for the samples in each regime we obtain the predictive scalings:
\begin{align}
    B_{pred, D}(\ell=1)\approx710\text{ G }\times P^{1/6},\label{eq:MDwarfD}\\
    B_{pred, ND}(\ell=1)\approx67\text{ G }\times P^{1/6}.\label{eq:MDwarfND}
\end{align}
It is difficult to say whether the spectropolarimetry measurements of these stars are more analogous to the surface values from our models or deeper. They do recover toroidal magnetic field strengths (from the limbs) which are zero at the surface of our models, due to the potential field boundary condition. In numerical models magnetic field amplitudes - in all components - decrease steeply as the outer boundary is approached (compare our scaling coefficients for $Lo_o$ and $Lo_V$, and see, for example, Fig.~4 from \citet{Yadav_2015}). Also, dipolarity tends to decrease towards the outer boundary - noting that in the bulk dipolarity also includes the toroidal contribution. However, the scalings for the dipole magnetic field strength that we obtain from our models have an offset that differs by an order of magnitude, both for $Lo_o(\ell=1)$ and $Lo_V(\ell=1)$, between dipolar and non-dipolar models. This makes both of them consistent with the coefficients of $\mathcal{C}_{\ell=1,D}=710\,$G and $\mathcal{C}_{\ell=1,ND}=67\,$G we find for these stars.

Interestingly, as with our numerical models, we do not find any star in our sample that indicates a bistability within a single system, i.e. even $B(\ell=1)$ measurements at different times, while showing some variability, do not jump between dipolar/non-dipolar regimes. However, both for our simulations and for these stars this may occur with longer run time/observation time.

We have excluded the very slow rotators (represented by hexagons in Fig.~\ref{fig:MDwarfs}) in the analysis outlined above. In the context of the study by \citet{Lehmann_2024} from which this data is taken, these stars have stronger magnetic fields than expected. This is because when considered purely as a function of rotation rate, magnetic field strength is found to decrease with increasing period, for $P_{rot}\gtrsim10\,$d \citep[see, e.g.][]{Reiners_2022}. However, as shown by Fig.~\ref{fig:MDwarfs}b and c, their magnetic field strengths are actually weaker with respect to their available driving power than the fast rotators. The derivation of the power-based scaling law explicitly requires the system to be rapidly rotating. Therefore, we do not expect it to correctly predict the magnetic field strengths of these slow rotators.

\subsection{Solar System Planets}\label{sec:PlanetsAppl}

   \begin{figure}[t!]
   \centering
   \includegraphics[width=0.8\hsize]{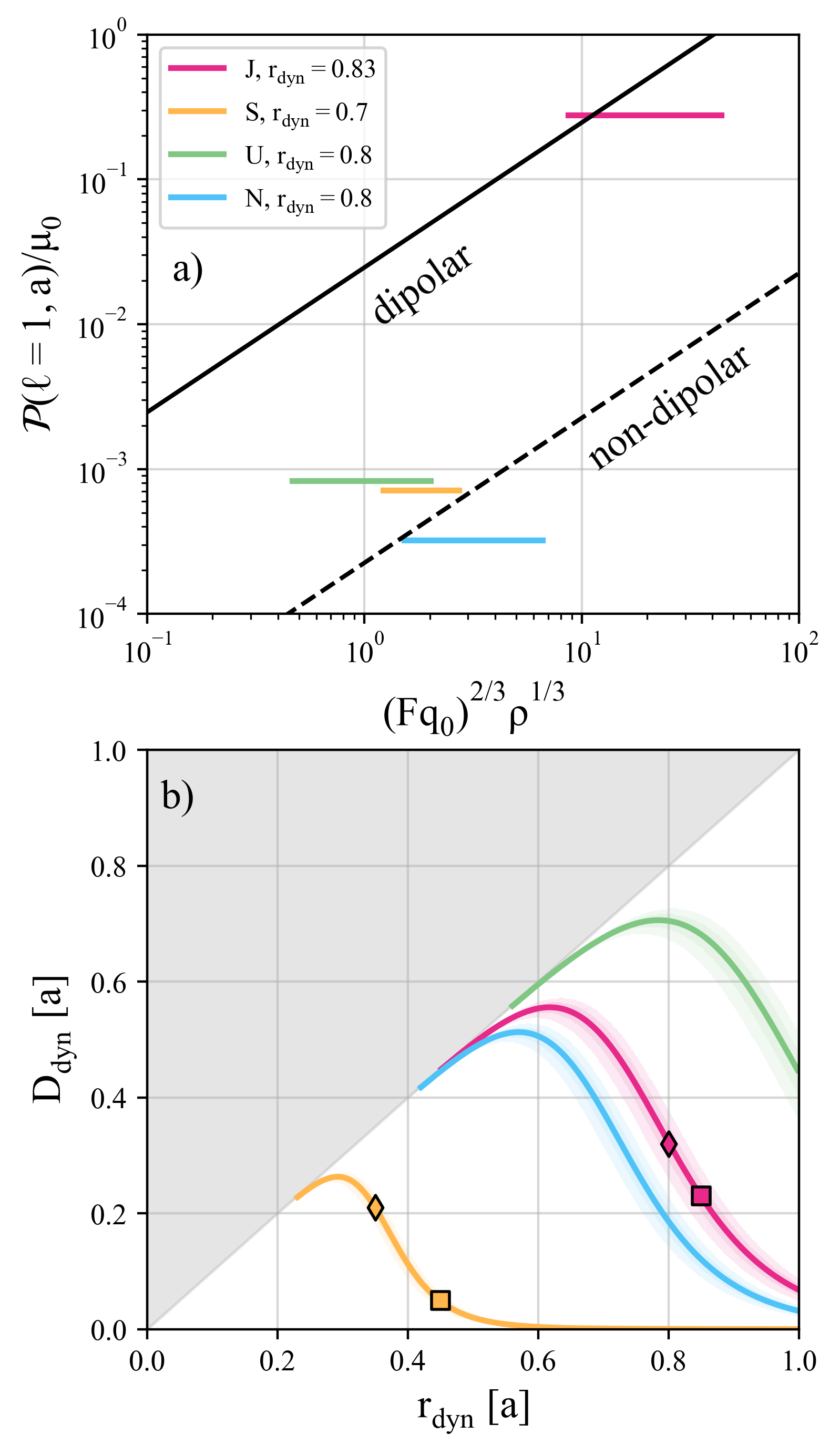}
   \includegraphics[width=0.83\hsize]{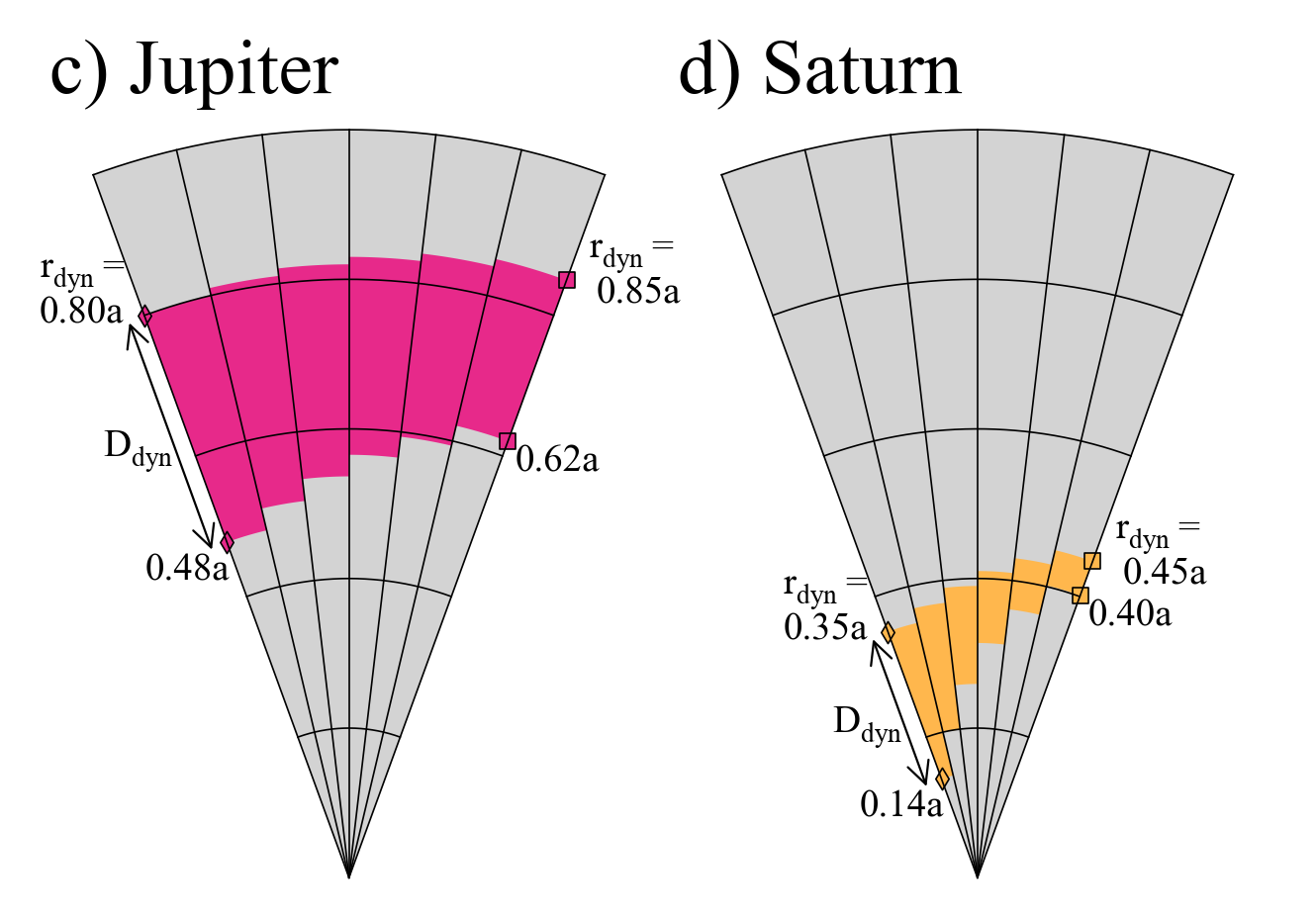}
      \caption{a) \textit{Dipole} magnetic field strength at the planetary surface, as a function of driving power, for the four giant planets in our solar system \citep[analogous to figures in e.g.][]{Christensen_2009}. The value we take for the dynamo radius for each planet is given in the legend and the horizontal extent of the power estimate comes from allowing for a range of $D_{dyn}$. The black solid/dashed line indicates the prediction for dipolar/non-dipolar magnetic fields. b) The $r_{dyn}$, $D_{dyn}$ parameter space in which the power in the planet's dipole matches the prediction within $20\%$. We use the dipolar scaling pre-factor for Jupiter and Saturn and the non-dipolar scaling pre-factor for Uranus and Neptune. Dynamo region geometry of c) Jupiter and d) Saturn, from the deepest realistic estimate (indicated by diamond markers) to shallowest estimate (square markers) indicated on panel b.}
         \label{fig:MagFieldScal_Planets}
   \end{figure}

One major snag when attempting to apply these scaling laws to planetary dynamos, is the fact planetary interiors could feature multiple stably stratified regions. This has become more apparent in the last decade with the measurements from Juno and Cassini at Jupiter and Saturn, respectively. It is now thought that both gas giants feature fuzzy cores \citep{Mankovich_2021, Militzer_2024, Helled_2024} as well as helium rain layers in shallower regions \citep[e.g.][]{Stevenson_1982, Markham_2024, Wulff_2025}. The ice giants are much less constrained due to lack of in-situ data, apart from Voyager 2's fly-by, but are also likely to host regions of compositional or thermal stratification, or phase changes \citep{Amoros_2024, Militzer_2024b, Gupta_2025, Morf_2025}.

The presence of these regions/boundaries complicates the application of the scaling laws. First, as even for Jupiter and Saturn their exact location is still poorly constrained, it is difficult to estimate the lower and upper bounds of the dynamo region. Second, the presence of stable stratification \textit{above} a dynamo region, but at a depth where electrical conductivity is still significant, has drastic effects on the magnetic field morphology. In particular, if differential rotation is strong in this layer, non-axisymmetric magnetic field components are filtered out effectively \citep[e.g.][]{Stevenson_1982, Christensen_2006b}. Thus, the field that is measured at the planetary surface can be very different to the field at the top of the dynamo region and cannot simply be extrapolated down as a potential field \citep{Wulff_2025}. Third, the case of a planet that has a truly nested dynamo system - i.e. multiple convective regions, separated by stable layers or phase changes, that all host dynamo action, not just magnetoconvection - has not been systematically explored.

Thus, with these caveats in mind, we test the accuracy of our scaling laws for the dipole magnetic field strength of the four giant planets in our solar system. Given that $\mathcal{P}_B(\ell,r)=\mathcal{P}_B(\ell,a)\,(a/r)^{2\ell+4}$, where $a$ is the planetary radius, the expression for $Lo(\ell=1)$ (see equation~\ref{eq:Lo}) at a planet's dynamo radius $r_{dyn}$ is:
\begin{equation}
    Lo_{dyn}(\ell=1)=\left(\frac{\mathcal{P}_B(\ell=1,a)\,(a/r_{dyn})^6}{\rho\mu_0\Omega^2 D_{dyn}^2}\right)^{1/2},
\end{equation}
where $\mathcal{P}_B(\ell=1,a)$ is the energy in its magnetic dipole at the planetary surface, and $D_{dyn}$ is the extent of the dynamo region (equal to $r_o-r_i$ in our numerical models).

The flux-based expression for $Ra^*_Q$ is:
\begin{equation}
    Ra^*_Q=\frac{1}{4\pi r_{dyn}(r_{dyn}-D_{dyn})}\frac{\alpha gQ_{adv}}{\rho c_P\Omega^3D_{dyn}^2}.
\end{equation}
We then substitute these definitions into the expression for the dipole Lorentz number $Lo_o(\ell=1)=c{Ra^*_Q}^{1/3}$, \EDIT{where we have followed the typically made assumption that $f_{ohm}=1$ in the planets, as we have no way of measuring this directly. See Appendix~\ref{sec:LoRoScale} for comments on the impact of this assumption. In our expression $c$ is the pre-factor from Table~\ref{tab:RaQFits}, where we test the dipolar pre-factor for Jupiter and Saturn and the non-dipolar one for Uranus and Neptune. This yields}:
\begin{align}
    \Bigg( & \frac{\mathcal{P}_B(\ell=1,a)\, a^6}{\rho\mu_0\Omega^2 D_{dyn}^2 r_{dyn}^6}\Bigg)^{1/2}\nonumber\\
    & =c \left(\frac{1}{4\pi r_{dyn}(r_{dyn}-D_{dyn})}\frac{\alpha gQ_{adv}}{\rho c_P\Omega^3D_{dyn}^2}\right)^{1/3},
\end{align}
and using the expression $Q_{adv}=4\pi r_{dyn}^2q_0$ we obtain
\begin{equation}\label{eq:PlanetScalings}
    \frac{\mathcal{P}_B(\ell=1,a)}{\mu_0}=c^2 \left(\frac{r_{dyn}^{10}D_{dyn}}{a^9(r_{dyn}-D_{dyn})}\frac{\alpha g q_0}{c_P}\right)^{2/3}\rho^{1/3}.
\end{equation}
We simplify the RHS of this equation and make it analogous to \citet{Christensen_2009}, by defining:
\begin{equation}
    F=\frac{r_{dyn}^{10}D_{dyn}}{a^9(r_{dyn}-D_{dyn})}\frac{\alpha g }{c_P}
\end{equation}
The values for $\alpha$, $g$, $c_P$ and $q_0$ used are given in Appendix~\ref{sec:PlanetData}. Note that the highest order dependence is on the dynamo radius. Thus, we mainly explore this dependence, though uncertainties in $\alpha$, $g$, $c_P$ and $q_0$ will also influence our results.

\begin{table}[h!]
\caption {Predictive scalings for the dipole magnetic field strength at the surface of a planet with a dynamo region spanning $r_{dyn}-D_{dyn}$ to $r_{dyn}$.\label{tab:PlanetScalings}} 
\centering
\begin{tabular}{ll}
\hline \hline 
Regime & Scaling Law \\
\hline \hline 
Dipolar & $\frac{\mathcal{P}_B(\ell=1,a)}{\mu_0}=(0.157 \pm 0.022)^2 (F q_0)^{2/3}\rho^{1/3}$ \\
Non-dipolar & $\frac{\mathcal{P}_B(\ell=1,a)}{\mu_0}=(0.015 \pm 0.003)^2 (F q_0)^{2/3}\rho^{1/3}$ \\
\hline
\end{tabular}
\end{table}

Fig.~\ref{fig:MagFieldScal_Planets}a shows how well the scaling laws given in Table~\ref{tab:PlanetScalings} agree with the measured dipole magnetic field energy for the solar system gas planets, with the two regimes indicated by the black lines. We make an estimate of the dynamo radius for each planet based on interior models and magnetic field measurements \citep{French_2012, Preising_2023, Wulff_2025}. The extent of each line on the x-axis is based on a range of values for $D_{dyn}$: $0.2-0.65\,a$ for Jupiter, $0.2-0.4\,a$ for Saturn, and $0.2-0.6\,a$ for Uranus and Neptune. Jupiter, Neptune and Uranus fall nicely within the range that we would expect from our scaling laws. However, Saturn is further from our predictions, based on our assumed dynamo radius and $D_{dyn}$ values.

If one makes no assumption of $r_{dyn}$ or $D_{dyn}$, one can map out where the scaling law holds in the $r_{dyn}$, $D_{dyn}$ parameter space, as illustrated in Fig.~\ref{fig:MagFieldScal_Planets}b. If our scaling law holds, this indicates that Jupiter and Neptune both have dynamo regions with similar trade-offs between dynamo region depth and extent. For Jupiter we have a better constraint on the dynamo radius and if $r_{dyn}=0.83\,R_J$, then this indicates that the dynamo region spans $\sim30\%$ of the planet, i.e. $0.55-0.83\,R_J$. Meanwhile, due to Uranus' extremely low heat flux, the scaling law implies that the dynamo region extends almost to the centre of the planet, and starts near to the surface. For Saturn, the scaling law suggests a deeper dynamo radius than what we assumed, based on its electrical conductivity profile.

We illustrate our favoured solutions for Jupiter and Saturn in Fig.~\ref{fig:MagFieldScal_Planets}c and d, respectively. Here, the solutions for Jupiter with $0.8R_J\leq r_{dyn}\leq0.85R_J$ \EDIT{and for Saturn with $0.35R_S\leq r_{dyn}\leq0.45R_S$,} are shown by highlighting the corresponding dynamo region extent, where the range is also indicated by the square and diamond markers on Fig.~\ref{fig:MagFieldScal_Planets}b.

\section{Conclusions}

We confirm the bistability of dynamo action in a wide range in $Ro_\ell$ (Inertia vs. Coriolis) space, where non-dipolar dynamos at even lower $Ro_\ell$ than presented here seem only limited by computational restrictions \EDIT{(i.e. they require lower Ekman numbers, which are more costly to simulate)}.

We find that the main difference in magnetic field amplitude lies in the power of the $\ell=1$ component and that across the remaining length-scales, power is distributed in a very similar way when comparing bistable systems.

We present scaling laws for both the volumetric and the surface total magnetic field amplitudes, as well as splitting them into their dipolar ($\ell=1$) and non-dipolar ($\ell>1$) parts. This allows for a better comparison with measurements of large-scale magnetic fields, as well as improved estimates for the dipole field strength which is an important controlling factor for stellar and planetary magnetospheres.

We show that these driving power-based scaling relations can also be formed in the framework of the system's characteristic Rossby number, according to the Coriolis-inertial-Archimedean (CIA) balance.

\EDIT{Naturally, the next step is to further investigate the dynamo mechanisms leading to the two regimes, and how they are linked to differences in zonal flow structures.}

\subsection{M dwarfs}

Bistability has previously been referred to when explaining the differences in the large-scale magnetic field structures inferred by ZDI for M-dwarfs \citep[e.g.][]{Morin_2011, Gastine_2013}. Here, we intentionally use the nomenclature "dipolar" and "non-dipolar", as opposed to "strong-field" and "weak-field" as our analysis showed that in this dual regime, the primary difference lies in the $\ell=1$ component of the magnetic field, while the strong- and weak-field nomenclature sometimes also refers to strong and weak dipolar bistable systems \citep[e.g.][]{Teed_2025}.

We have built upon the scaling law which predicts the total magnetic field amplitude \citep[Eq.~\ref{eq:MDwarfTot}][]{Christensen_2009}, by expanding to predictions of the magnetic dipole, Eq.~\ref{eq:MDwarfD} and \ref{eq:MDwarfND}, for dipolar and non-dipolar systems, respectively. We show that these differ by an order of magnitude between the two regimes. If one has an estimate of the differential rotation occurring in the star, this can help to distinguish which regime it is likely to be in, and which scaling law one should apply \citep{Gastine_2014, Zaire_2022}. If it is a rapid rotator, and the differential rotation is strong and solar-like, then it is more likely to be in the non-dipolar regime. If the differential rotation is weak, then there is likely a strongly dipolar magnetic field braking the zonal flow and ones should apply the dipolar scaling law.

\citet{Donati_2025} investigate the effect of adding more information into the ZDI analysis, using not just Stokes V, but also Stokes IV and Stokes IVQU profiles. They find that the inferred dipole strength is stronger compared to the results when only using Stokes V signatures. However, they express this as the polar field strength, and only give the axisymmetric dipolarity. This can differ quite significantly from the total dipolarity for non-dipolar magnetic field morphologies where the dipole has less preference for being aligned with the rotation axis. Nonetheless, their analysis provides a note of caution that our M dwarf dipole magnetic field strength predictions are based on retrievals from Stokes V and may need to be adapted as measured estimates of the large-scale magnetic field components of low-mass stars are improved.

Our models miss the very small-scale magnetic field structures. We do not analyse the scale-filtering, and comparison between models and observational methods in more detail in this work \citep[see][]{Yadav_2015, Lehmann_2018}. However, \citet{Lang_2014} find that while adding small-scale field increases the surface flux, the large-scale open flux that governs the stellar magnetosphere remains unaffected. 

Thus, our predictions are useful when considering the magnetic environment of exoplanets, which are primarily shaped by the host star's large-scale field \citep{Vidotto_2013, Cohen_2014}. In particular, the M dwarfs with non-dipolar magnetic fields may make the best host stars for habitable exoplanets, as \citet{Vidotto_2014b} find that less axisymmetric stellar magnetic fields help to shield the planets from galactic cosmic rays more efficiently. 

\subsection{Solar System Planets}

Previous suggestions for the reason behind the non-dipolar nature of the ice giants' magnetic fields have included the presence of a stably stratified layer below the dynamo region \citep{Stanley_2004, Stanley_2006}, reducing the dynamo region to a thinner layer, or the dominance of inertia over rotational effects in their internal dynamics \citep{Soderlund_2012, Soderlund_2013}. However, we show that the non-dipolar regime has a wide range of overlap with the dipolar regime in $Ro_\ell$ space and that neither of these ideas need to be invoked to explain their weak $\ell=1$ magnetic field components. Our scaling laws do indicate a deeper dynamo region on Uranus, compared to Neptune, due to its extremely low heat flux. However, we observed more variation of the energy in the dipole in our non-dipolar models, due to its subservience in these systems. Thus, the energy in the magnetic dipole of Uranus/Neptune may currently be in a stronger/weaker phase, respectively. In general, these systems are better characterised by the energy in the non-dipolar large-scale magnetic field which we hope may be measured by future missions.

Our dipolar scaling law applied to Saturn indicates that its dynamo operates much deeper than one would infer from electrical conductivity profiles alone \citep[as also suggested by][]{Christensen_2018}. This is also the case in Jupiter. \citet{Wulff_2025} show that estimating the magnetic Reynolds number (an indicator of dynamo action) as a function of radius predicts a dynamo radius of around $0.91\,R_J$ \citep[electrical conductivity $\sigma\approx10^5\,S/m$ at this depth;][]{French_2012}. A similar electrical conductivity is reached at $0.7-0.75\,R_S$ in Saturn \citep{Preising_2023}.

However, the magnetic field spectrum at the Jovian surface indicates $r_{dyn}\approx0.83\,R_J$ \citep[obtained by carrying out a Lowes analysis,][]{Connerney_2022, Wulff_2025}. This mismatch is due to the presence of a stably stratified helium rain layer between $\sim0.83\,R_J$ to around $0.87-0.9\,R_J$. Such a layer is also present in Saturn and is widely held responsible for the extremely axisymmetric nature of its magnetic field at the surface \citep{Stevenson_1982, Cao_2020, Yadav_2022}. As this is a much more pronounced effect than at Jupiter, the helium rain layer is thought to be much thicker inside Saturn, which is also endorsed by stability analysis in the two planets \citep{Markham_2024}.

A Lowes radius analysis cannot be easily performed at Saturn due to the irregular zig-zags in its surface spectrum - markers of the magnetic field filtering process occurring in the stable layer. However, the distribution of magnetic energy in its odd spectral components (excluding $\ell=5$) closely mirrors that of the Earth, which has a dynamo radius of $\sim0.55\,R_E$. This indicates that the dynamo radius is deeper than the $0.7-0.75\,R_S$ implied by conductivity alone.

Thus, both the axisymmetric nature of the Saturnian magnetic field, and the gradient of the magnetic power spectrum are consistent with the deeper dynamo radius ($r_{dyn}\approx0.45\,R_S$) suggested by our scaling law analysis. This is at odds with ring seismology analysis which suggests the inner $\sim60\%$ of Saturn is stably stratified \citep{Mankovich_2021}. If we take $r_{dyn}=0.7\,R_S$ and $D_{dyn}=0.1\,R_S$, as this would imply, then this would require $\alpha gq_0/c_P\approx2.18\times10^{-17}$ for the scaling law to hold, which is $0.00052\%$ of our estimate, based on Saturnian interior models. It is unlikely that our estimates of $\alpha$, $g$, $q_0$ and $c_P$ are so far off. The $\mathcal{P}_B(\ell=1,a)\sim r_{dyn}^6$ dependence in eq.~\ref{eq:PlanetScalings} means that small adjustments in the estimate of $r_{dyn}$ are much more effective in finding feasible solutions that obey our scaling laws.

As alluded to in section~\ref{sec:PlanetsAppl}, the filtering process occurring in the stable helium rain layer in Saturn also means that we must take caution when testing the scaling law on the dipole field strength measured at the surface. The assumption that the field behaves as a potential in this layer, and thus that $\mathcal{P}_B(\ell,r)=\mathcal{P}_B(\ell,a)\,(a/r)^{2\ell+4}$ holds does not reflect the more complex effect of the stable stratification, in the presence of strong differential rotation.

Furthermore, the mechanical, thermal and magnetic boundary conditions used in our suite of fully conducing, fully convective dynamo simulations do not reflect the complexity of the interfaces with the regions above and below the dynamo regions in the giant planets. However, \citet{Yadav_2013b} showed that the scaling of $Lo_V$ held for a variety of numerical models, including those with different boundary conditions and varying background density and conductivity profiles. This gives us confidence that our simpler models can be extended to planetary interiors, within reasonable uncertainties.

\subsection{Exoplanets}

When it comes to predicting the magnetic field strength morphology of exoplanets the main motivations, currently, are either to investigate the induction that may take place in the thermally ionised atmospheres of Hot Jupiters, or to identify likely candidates for magnetic field detection. In both cases it is always assumed that the planet has a dipolar magnetic field morphology \citep[e.g.][]{Batygin_2010, Dietrich_2022, Kilmetis_2024, Elias_Lopez_2025}.

\citet{Christensen_2006, Christensen_2010} suggest a factor of around 3.5 for scaling up the magnetic field strength at the top of a dynamo region to the interior field strength. We obtain a similar ratio for the scalings of total magnetic field in dipolar dynamo cases as comparing the pre-factors given in Table~\ref{tab:RaQFits} yields $0.646/0.168\approx3.85$. Thus, our results indicate that using the established [CA06]/[C09] scaling laws to first infer internal magnetic field strength and then scale down using this factor to estimate total surface magnetic field strength should work well \textit{for dipolar magnetic field morphologies}. However, if one either has no measurement of the total magnetic field at the surface, or if one wants to estimate the dipole field strength specifically - as it is the most relevant for many space physics processes - one must consider the scalings for $B(\ell=1)$ specifically. This study highlights that, in order to do this, one must take into account that the magnetic field morphology may be non-dipolar, and that this makes up an order of magnitude difference. Further, it is then much more likely that the dipole component is not aligned with the rotation axis and is fluctuating more in time.

\begin{acknowledgements}
We thank Denis Shulyak for acting as consultant on magnetic field measurements of M Dwarfs. PNW and HC thank the NASA Juno project and the NASA CDAP grant No. 80NSSC23K0511 for support. HC also acknowledges the support from the Sloan Research Fellowship.
\end{acknowledgements}

\appendix

\section{\EDIT{Initialisation of Simulations}}\label{sec:Init}

   \begin{figure*}[h!]
   \centering
   \includegraphics[width=0.5\hsize]{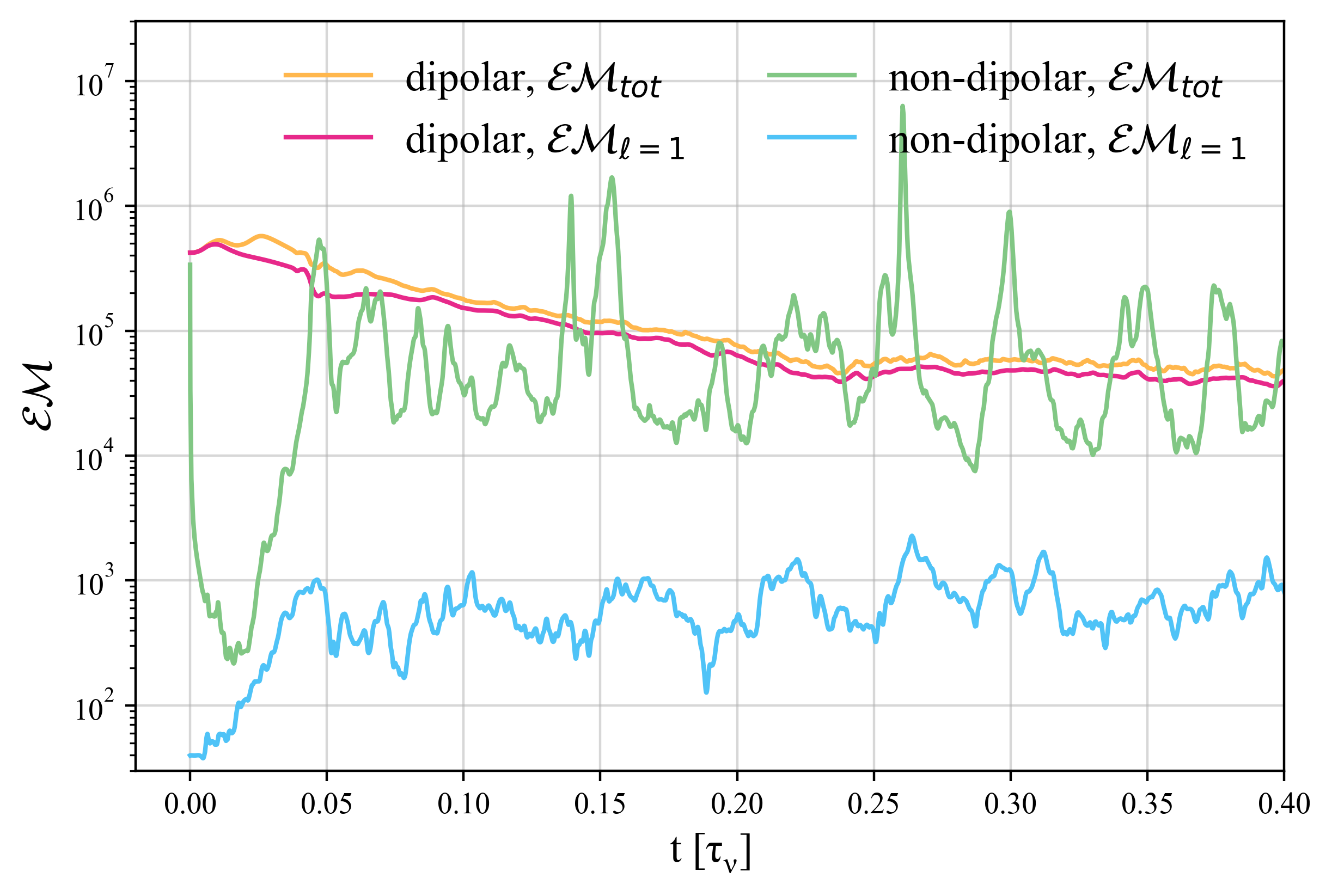}
\caption{Time series for two simulations with the same control parameters. The orange (pink) curve represents the evolution of the total (dipole) energy for the model that has a steady state dominated by the dipole. Green (blue) are the equivalents for the model in the non-dipolar regime.
\label{fig:dipTseries}}
\end{figure*}

\EDIT{We initialise the first dipole dominated models with an axial dipole with amplitude $\Lambda\sim5$ at $r=r_i$, and the non-dipolar models with a weak $\Lambda\sim0.1$ field across all modes ($\ell$ and $m$) randomly. The evolution of the total magnetic energy and the energy in the dipole (axial and equatorial) for one pair of models is shown in Figure~\ref{fig:dipTseries}. These both have control parameters $Pm=1$, $E=10^{-4}$ and $Ra=10^{7}$.}

\begin{table}[h!]
\section{Magnetic Field Scaling with Rossby number}\label{sec:LoRoScale}
\caption {Coefficients obtained when fitting \EDIT{$Lo\sim{Ra_Q^*}^{1/3}$,} $Lo\sim{Ro_c}^{5/3}$, $Lo\sim Ro^{5/6}$, and $Lo\sim Ro_\ell^{5/3}$. \label{tab:AllFits}} 
\centering
\begin{tabular}{lcccc}
\hline\hline             
 & \multicolumn{4}{c}{Pre-factor}\\
Scaling & \EDIT{${Ra_Q^*}^{1/3}$} & ${Ro_c}^{5/3}$ & ${Ro^\prime}^{5/6}$ & $Ro_\ell^{5/3}$ \\
\hline\hline
$Lo_o$ Dipolar & $0.121\pm0.016$ & $0.027 \pm 0.003$ & $0.138 \pm 0.019$ & $0.236 \pm 0.067$ \\
$Lo_o$ Non-dipolar & $0.047\pm0.006$ & $0.009 \pm 0.001$ & $0.049 \pm 0.006$ & $0.068 \pm 0.010$ \\
$Lo_o(\ell=1)$ Dipolar & $0.113\pm0.015$ & $0.025 \pm 0.003$ & $0.129 \pm 0.019$ & $0.221 \pm 0.063$ \\
$Lo_o(\ell=1)$ Non-dipolar & $0.009\pm0.002$ & $0.002 \pm 0.000$ & $0.010 \pm 0.002$ & $0.013 \pm 0.003$ \\
$Lo_o(\ell>1)$ Universal & $0.045\pm0.004$ &$0.009 \pm 0.001$ & $0.048 \pm 0.004$ & $0.069 \pm 0.010$ \\
\hline
$Lo_V$ Dipolar & $0.465\pm0.052$ &$0.102 \pm 0.011$ & $0.531 \pm 0.066$ & $0.898 \pm 0.271$ \\
$Lo_V$ Non-dipolar & $0.294\pm0.041$ & $0.059 \pm 0.008$ & $0.309 \pm 0.041$ & $0.419 \pm 0.084$ \\
$Lo_V(\ell=1)$ Dipolar & $0.402\pm0.044$ & $0.088 \pm 0.009$ & $0.459 \pm 0.057$ & $0.769 \pm 0.245$ \\
$Lo_V(\ell=1)$ Non-dipolar & $0.042\pm0.011$ & $0.009 \pm 0.002$ & $0.044 \pm 0.012$ & $0.058 \pm 0.021$ \\
$Lo_V(\ell>1)$ Universal & $0.274\pm0.032$ & $0.056 \pm 0.006$ & $0.295 \pm 0.031$ & $0.421 \pm 0.065$ \\
\hline
\end{tabular}
\end{table}

\EDIT{In Table~\ref{tab:AllFits} we present the coefficients for the fitting of Lorentz numbers with ${Ra_Q^*}^{1/3}$ (i.e. omitting the factor of $f_{ohm}^{1/2}$). Comparing these values with those given in Table~\ref{tab:RaQFits} highlights the implications of assuming $f_{ohm}=1$ for planets: if we take values of the ohmic dissipation fraction that are more similar to our models, we would infer significantly thicker dynamo regions for Jupiter and Saturn. As an example, for a dynamo radius of $0.8R_J$ for Jupiter, we would estimate a lower dynamo boundary of $\sim0.29R_J$, compared with $0.48R_J$ obtained with the scaling that assumes that in Jupiter $f_{ohm}=1$ (see Fig.~\ref{fig:MagFieldScal_Planets}c). }

\begin{figure*}[h!]
   \centering
   \includegraphics[width=\hsize]{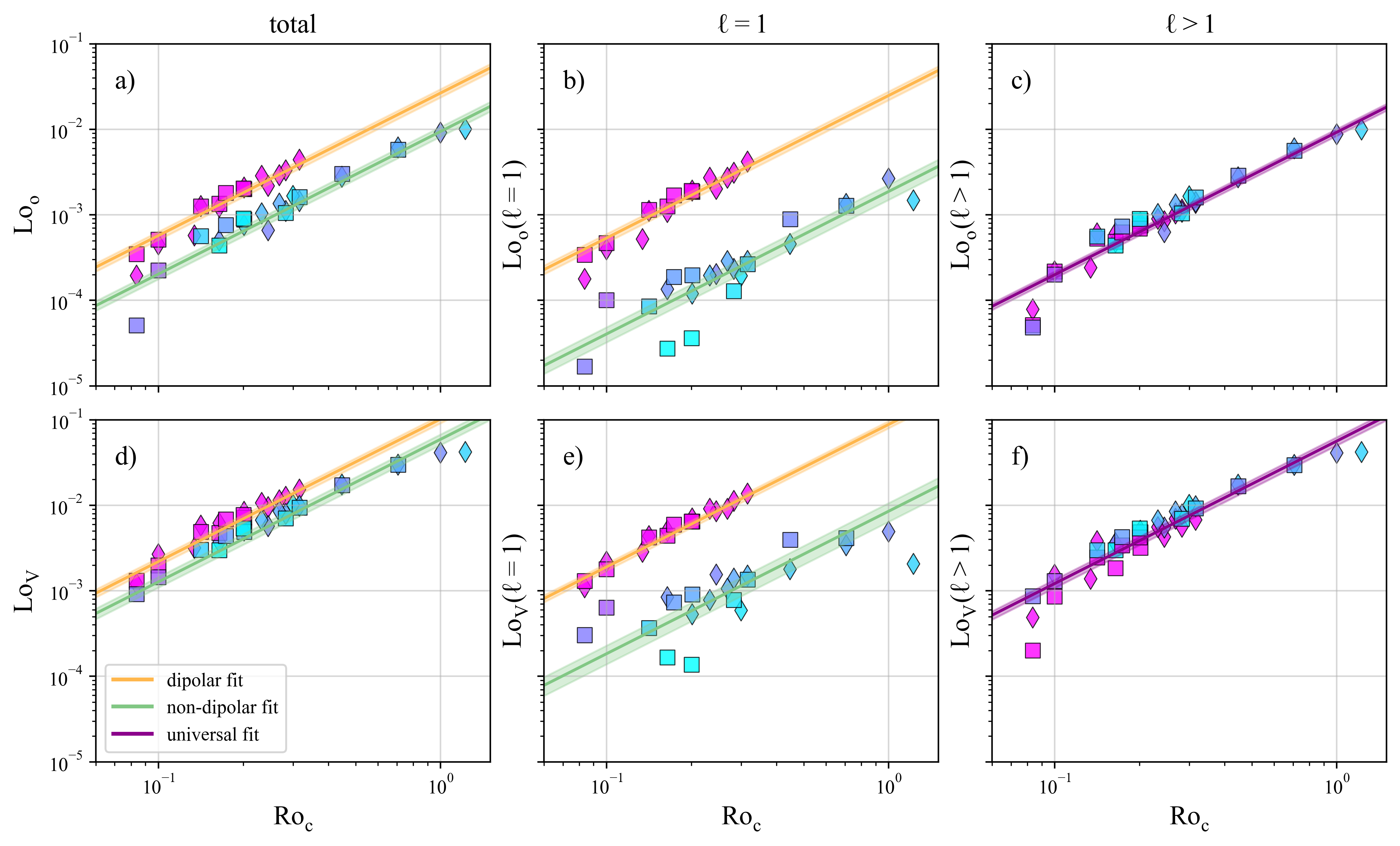}
\caption{The equivalent of Figure~\ref{fig:LoRaQ}, but with convective Rossby number $Ro_c$ on the x-axis and with $Lo$ rather than $Lo/f_{ohm}^{1/2}$ on the y-axis. See Table~\ref{tab:AllFits} for the parameters of the lines of best fit.
\label{fig:LoRoc}}
\end{figure*}

\EDIT{Table~\ref{tab:AllFits} also gives the coefficients for the fitting $Lo$ with the various Rossby numbers, with their respective exponents. The equivalent of Figure~\ref{fig:LoRaQ}, but with the input parameter $Ro_c$ on the x-axis is shown in Fig.~\ref{fig:LoRoc}.}

\section{M Dwarf Data}\label{sec:MDwarfData}

\EDIT{Here we provide the data used for our investigation of M Dwarf magnetic field amplitudes. Table~\ref{tab:MDwarfs} gives both the data, as well as the sources (see caption for abbreviations).}
\begin{table}[h!]
\caption {Stellar properties. In the references M10 refers to \cite{Morin_2010}, B24 to \cite{Bellotti_2024}, L24 to \cite{Lehmann_2024}, M08b to \cite{Morin_2008b}, M08a to \cite{Morin_2008a}, R22 to \cite{Reiners_2022}, and S17 to \cite{Shulyak_2017}.  \label{tab:MDwarfs}} 
\centering
\begin{tabular}{l|llllllll}
\hline\hline
Star & $M$ & $L$ & $R$ & $P_{rot}$ & $B(\ell=1)$ & $B$ & ref. & ref. \\
 & [$M_\odot$] & [$L_\odot$] & [$R_\odot$] & [d] & [G] & [G] & (Stokes V) & (Stokes I) \\
\hline\hline 
GJ 1111 (2007)	& 0.1	& $7.09\times10^{-4}$	& 0.11	& 0.46	& 92.3	& 2680	& M10 & R22 \\
GJ 1111 (2008)	& 0.1	& $7.09\times10^{-4}$	& 0.11	& 0.46	& 45.3	& 2680	& M10  & R22 \\
GJ 1111 (2009)	& 0.1	& $7.09\times10^{-4}$	& 0.11	& 0.46	& 53.5	& 2680	& M10  & R22 \\
Gl 412 B (2006)	& 0.1	& $1.05\times10^{-3}$	& 0.12	& 0.78	& 904	& 6880	& M10  & R22 \\
Gl 412 B (2007)	& 0.1	& $1.05\times10^{-3}$	& 0.12	& 0.78	& 1160	& 6880	& M10  & R22 \\
Gl 412 B (2008)	& 0.1	& $1.05\times10^{-3}$	& 0.12	& 0.78	& 1110	& 6880	& M10  & R22 \\
Gl 412 B (2009)	& 0.1	& $1.05\times10^{-3}$	& 0.12	& 0.78	& 1590	& 6880	& M10  & R22 \\
Gl 406 (2019a)	& 0.1	& $7.41\times10^{-4}$	& 0.12	& 2.7	& 348	& 2600	& B24& S17 \\
Gl 406 (2019b2020a)	& 0.1	& $7.41\times10^{-4}$	& 0.12	& 2.7	& 319	& 2600	& B24	& S17 \\
Gl 406 (2020b2021a)	& 0.1	& $7.41\times10^{-4}$	& 0.12	& 2.7	& 482	& 2600	& B24 & S17 \\
Gl 406 (2021b2022a)	& 0.1	& $7.41\times10^{-4}$	& 0.12	& 2.7	& 415	& 2600	& B24 & S17 \\
GJ 1245 B (2006)	& 0.12	& $1.74\times10^{-3}$	& 0.14	& 0.71	& 122	& 3400	& M10	& R22 \\
GJ 1245 B (2008)	& 0.12	& $1.74\times10^{-3}$	& 0.14	& 0.71	& 38.2	& 3400	& M10& R22 \\
GJ 1286 (2020)	& 0.12	& $1.40\times10^{-3}$	& 0.142	& 178	& 103	& 910	& L24	& R22 \\
GJ 1286 (2021)	& 0.12	& $1.40\times10^{-3}$	& 0.142	& 178	& 67.3	& 910	& L24	& R22 \\
GJ 1156 (2007)	& 0.14	& $2.16\times10^{-3}$	& 0.16	& 0.49	& 35.3	& 3970	& M10& R22 \\
GJ 1156 (2008)	& 0.14	& $2.16\times10^{-3}$	& 0.16	& 0.49	& 74.9	& 3970	& M10	& R22 \\
GJ 1156 (2009)	& 0.14	& $2.16\times10^{-3}$	& 0.16	& 0.49	& 71.4	& 3970	& M10	& R22 \\
Gl 905 (2019)	& 0.15	& $2.18\times10^{-3}$	& 0.165	& 114.3	& 124	& 410	& L24	& R22 \\
Gl 905 (2020)	& 0.15	& $2.18\times10^{-3}$	& 0.165	& 114.3	& 80.2	& 410	& L24	& R22 \\
Gl 905 (2021)	& 0.15	& $2.18\times10^{-3}$	& 0.165	& 114.3	& 63.6	& 410	& L24	& R22 \\
GJ 1151 (2019)	& 0.17	& $3.42\times10^{-3}$	& 0.193	& 175.6	& 22.7	& 680	& L24	& R22 \\
GJ 1151 (2020)	& 0.17	& $3.42\times10^{-3}$	& 0.193	& 175.6	& 31.9	& 680	& L24	& R22 \\
GJ 1151 (2021)	& 0.17	& $3.42\times10^{-3}$	& 0.193	& 175.6	& 61.7	& 680	& L24	& R22 \\
GJ 1289 (2019)	& 0.21	& $5.38\times10^{-3}$	& 0.233	& 73.66	& 79.2	& 1360	& L24	& R22 \\
GJ 1289 (2020)	& 0.21	& $5.38\times10^{-3}$	& 0.233	& 73.66	& 194	& 1360	& L24 & R22 \\
GJ 1289 (2021)	& 0.21	& $5.38\times10^{-3}$	& 0.233	& 73.66	& 199	& 1360	& L24	& R22 \\
EQ Peg B (2006)	& 0.25	& $7.70\times10^{-3}$	& 0.25	& 0.4	& 379	& 4200	& M08b	& R22 \\
V374 Peg (2005)	& 0.28	& $9.47\times10^{-3}$	& 0.28	& 0.45	& 588	& 3630	& M08a	& R22 \\
V374 Peg (2006)	& 0.28	& $9.47\times10^{-3}$	& 0.28	& 0.45	& 491	& 3630	& M08a	& R22 \\
EV Lac (2006)	& 0.32	& $1.05\times10^{-2}$	& 0.3	& 4.37	& 447	& 4320	& M08b	& R22 \\
EV Lac (2007)	& 0.32	& $1.05\times10^{-2}$	& 0.3	& 4.37	& 420	& 4320	&M08b	& R22 \\
EV Lac (2019b)	& 0.32	& $1.05\times10^{-2}$	& 0.3	& 4.37	& 63.1	& 4320	& B24	& R22 \\
EV Lac (2020b2021a)	& 0.32	& $1.05\times10^{-2}$	& 0.3	& 4.37	& 289	& 4320	& B24	& R22 \\
EV Lac (2021b)	& 0.32	& $1.05\times10^{-2}$	& 0.3	& 4.37	& 254	& 4320	& B24	& R22 \\
Gl 285 (2007)	& 0.32	& $7.27\times10^{-3}$	& 0.29	& 2.77	& 540	& 4540	& M08b	& R22 \\
Gl 285 (2008)	& 0.32	& $7.27\times10^{-3}$	& 0.29	& 2.77	& 514	& 4540	& M08b	& R22 \\
\hline
\end{tabular}
\end{table}
\newpage
\section{Solar System Gas Planets Data}\label{sec:PlanetData}

\EDIT{Data used in Section~\ref{sec:PlanetsAppl} is provided here in Table~\ref{tab:PlanBfields}.}

\begin{table}[h!]
\caption {Parameters used for the approximation of $Lo(\ell=1)$ at the dynamo surface of the solar system planets. \label{tab:PlanBfields}} 
\centering
\begin{tabular}{l|lllllll}
\hline\hline
Planet & $\mathcal{P}_B(\ell=1,a)$ & $a$ & $\alpha$ & $g$ & $q_0$ & $c_P$ & $\rho$ \\
 & [$nT^2$] & [$m$] & [$K^{-1}$] & [$m/s^2$] & [$W/m^2$] & [$JK^{-1}kg^{-1}$] & [$kg/m^3$] \\
\hline\hline 
Jupiter & $3.49\times10^{11}$ & $6.9911\times10^7$ & $1.2\times10^{-5}$ & 25.9 & 7.5 & $1.4\times10^4$ & 2000 \\
Saturn & $8.94\times10^8$ & $5.8232\times10^7$ & $2.5\times10^{-5}$ & 11 & 2 & $1.7\times10^4$ & 1500 \\
Uranus & $1.04\times10^9$ & $2.5362\times10^7$ & $4.4\times10^{-5}$ & 9 & 0.089 & $4.2\times10^3$ & 1750 \\
Neptune & $4.06\times10^8$ & $2.4622\times10^7$ & $4.4\times10^{-5}$ & 11.3 & 0.433 & $4.2\times10^3$ & 1750 \\
\hline
\end{tabular}
\end{table}

\begin{table}[h!]
\section{Model Parameters and Diagnostics}\label[sec:Models]
\centering
\caption{Models and their relevant control and diagnostic parameters. Critical Rayleigh numbers are $Ra_c=6.51\times10^5$ for $E=10^{-4}$, $Ra_c=2.68\times10^6$ for $E=3\times10^{-5}$, and $Ra_c=1.01\times10^7$ for $E=10^{-5}$. All simulations are Boussinesq, have aspect ratio $r_i/r_o=0.35$, $Pr=1$, and the same boundary conditions.}
\begin{tabular}{llcc|llllllll}
$E$ & $Ra$ & $Pm$ & $Ro_c$ & $Nu$ & $Ro^\prime$ & $\mathcal{L}_U$ & $f_{dip,r_o}$ & $f_{dip,V}$ & $\Lambda_o$ & $\Lambda_V$ & \EDIT{$f_{ohm}$} \\
\hline\hline
$10^{-4}$ & $6\times10^6$ & 1 & 0.24 & 2.43 & 0.008 & 0.185 & 0.847 & 0.798 & 0.047 & 0.906 & 0.438 \\
$10^{-4}$ & $6\times10^6$ & 1 & 0.24 & 2.26 & 0.008 & 0.182 & 0.093 & 0.075 & 0.004 & 0.325 & 0.29 \\
\hline
$10^{-4}$ & $8\times10^6$ & 1 & 0.28 & 3.11 & 0.01 & 0.177 & 0.879 & 0.8 & 0.11 & 1.578 & 0.473 \\
$10^{-4}$ & $8\times10^6$ & 1 & 0.28 & 3.08 & 0.011 & 0.176 & 0.049 & 0.033 & 0.012 & 0.627 & 0.322 \\
$10^{-4}$ & $8\times10^6$ & 0.5 & 0.28 & 2.9 & 0.011 & 0.174 & 0.015 & 0.013 & 0.006 & 0.247 & 0.308 \\
\hline
$10^{-4}$ & $10^7$ & 1 & 0.32 & 3.79 & 0.013 & 0.171 & 0.903 & 0.804 & 0.195 & 2.33 & 0.492 \\
$10^{-4}$ & $10^7$ & 1 & 0.32 & 3.82 & 0.014 & 0.177 & 0.039 & 0.023 & 0.021 & 0.973 & 0.35 \\
$10^{-4}$ & $10^7$ &  0.5 &  0.32 & 3.72 & 0.014 & 0.176 & 0.027 & 0.021 & 0.013 & 0.438 & 0.352 \\
\hline
$10^{-4}$ & $2\times10^7$ & 1 & 0.45 & 6.91 & 0.025 & 0.174 & 0.028 & 0.011 & 0.078 & 3.62 & 0.4 \\
$10^{-4}$ & $2\times10^7$ &  0.5 &  0.45 & 6.89 & 0.025 & 0.173 & 0.085 & 0.052 & 0.046 & 1.473 & 0.401 \\
\hline
$10^{-4}$ & $5\times10^7$ & 1 & 0.71 & 12.81 & 0.048 & 0.163 & 0.046 & 0.012 & 0.377 & 8.9 & 0.409 \\
$10^{-4}$ & $5\times10^7$ &  0.5 &  0.71 & 12.63 & 0.047 & 0.161 & 0.048 & 0.019 & 0.168 & 4.443 & 0.412 \\
\hline
$10^{-4}$ & $10^8$ & 1 & 1.0 & 18.68 & 0.072 & 0.151 & 0.083 & 0.014 & 0.836 & 17.013 & 0.427 \\
\hline
$10^{-4}$ & $1.5\times10^8$ & 1 & 1.22 & 23.22 & 0.102 & 0.16 & 0.023 & 0.002 & 1.011 & 17.627 & 0.385 \\
\hline
$3\times10^{-5}$ & $2\times10^7$ & 1 & 0.13 & 1.89 & 0.002 & 0.142 & 0.823 & 0.809 & 0.011 & 0.335 & 0.443 \\
\hline
$3\times10^{-5}$ & $3\times10^7$ & 1 & 0.16 & 2.87 & 0.004 & 0.128 & 0.758 & 0.662 & 0.051 & 1.273 & 0.563 \\
$3\times10^{-5}$ & $3\times10^7$ & 1 & 0.16 & 2.97 & 0.004 & 0.122 & 0.073 & 0.053 & 0.008 & 0.458 & 0.365 \\
$3\times10^{-5}$ & $3\times10^7$ &  0.5 &  0.16 & 2.79 & 0.004 & 0.116 & 0.87 & 0.85 & 0.03 & 0.381 & 0.408 \\
$3\times10^{-5}$ & $3\times10^7$ &  0.5 &  0.16 & 2.77 & 0.004 & 0.119 & 0.005 & 0.004 & 0.003 & 0.149 & 0.311 \\
\hline
$3\times10^{-5}$ & $4.5\times10^7$ & 1 & 0.2 & 4.43 & 0.006 & 0.118 & 0.843 & 0.696 & 0.145 & 2.347 & 0.565 \\
$3\times10^{-5}$ & $4.5\times10^7$ & 1 & 0.2 & 4.44 & 0.007 & 0.122 & 0.025 & 0.011 & 0.019 & 0.856 & 0.384 \\
$3\times10^{-5}$ & $4.5\times10^7$ &  0.5 &  0.2 & 4.13 & 0.006 & 0.115 & 0.857 & 0.803 & 0.067 & 0.866 & 0.484 \\
$3\times10^{-5}$ & $4.5\times10^7$ &  0.5 &  0.2 & 4.36 & 0.007 & 0.123 & 0.05 & 0.035 & 0.013 & 0.404 & 0.385 \\
\hline
$3\times10^{-5}$ & $6\times10^7$ & 1 & 0.23 & 5.69 & 0.007 & 0.116 & 0.901 & 0.729 & 0.272 & 3.785 & 0.602 \\
$3\times10^{-5}$ & $6\times10^7$ & 1 & 0.23 & 5.79 & 0.009 & 0.125 & 0.036 & 0.014 & 0.036 & 1.499 & 0.432 \\ 
\hline
$3\times10^{-5}$ & $8\times10^7$ & 1 & 0.27 & 7.5 & 0.01 & 0.117 & 0.873 & 0.631 & 0.282 & 4.359 & 0.555 \\
$3\times10^{-5}$ & $8\times10^7$ & 1 & 0.27 & 7.37 & 0.011 & 0.125 & 0.047 & 0.015 & 0.061 & 2.426 & 0.465 \\
\hline
$3\times10^{-5}$ & $10^8$ & 1 & 0.3 & 8.99 & 0.013 & 0.125 & 0.015 & 0.003 & 0.092 & 3.411 & 0.485 \\
\hline
$10^{-5}$ & $7\times10^7$ & 1 & 0.08 & 1.63 & 0.001 & 0.115 & 0.838 & 0.835 & 0.004 & 0.144 & 0.416 \\
$10^{-5}$ & $7\times10^7$ &  0.5 &  0.08 & 1.64 & 0.001 & 0.12 & 0.106 & 0.103 & 0.0 & 0.042 & 0.353 \\
\hline
$10^{-5}$ & $10^8$ & 1 & 0.1 & 2.51 & 0.002 & 0.097 & 0.762 & 0.669 & 0.02 & 0.717 & 0.58 \\
$10^{-5}$ & $10^8$ & 0.5 & 0.1 & 2.4 & 0.002 & 0.085 & 0.821 & 0.812 & 0.013 & 0.196 & 0.406 \\
$10^{-5}$ & $10^8$ &  0.5 &  0.1 & 2.48 & 0.002 & 0.089 & 0.202 & 0.192 & 0.003 & 0.105 & 0.36 \\
\hline
$10^{-5}$ & $2\times10^8$ &  1 & 0.14 & 5.96 & 0.003 & 0.088 & 0.765 & 0.562 & 0.16 & 3.345 & 0.676 \\
$10^{-5}$ & $2\times10^8$ &  0.5 &  0.14 & 5.43 & 0.003 & 0.083 & 0.824 & 0.747 & 0.079 & 1.193 & 0.585 \\
$10^{-5}$ & $2\times10^8$ &  0.5 &  0.14 &  5.39 & 0.004 & 0.091 & 0.024 & 0.015 & 0.016 & 0.457 & 0.423\\
\hline
$10^{-5}$ & $3\times10^8$ &  0.5 &  0.17 & 8.08 & 0.005 & 0.083 & 0.881 & 0.754 & 0.164 & 2.351 & 0.633 \\
$10^{-5}$ & $3\times10^8$ &  0.5 & 0.17 & 8.15 & 0.006 & 0.094 & 0.06 & 0.028 & 0.029 & 0.957 & 0.507 \\
\hline
$10^{-5}$ & $4\times10^8$ &  0.5 &  0.2 & 10.52 & 0.006 & 0.085 & 0.884 & 0.699 & 0.207 & 2.981 & 0.634 \\
$10^{-5}$ & $4\times10^8$ &  0.5 &  0.2 & 10.68 & 0.007 & 0.091 & 0.002 & 0.001 & 0.041 & 1.436 & 0.522 \\
\hline
\end{tabular}\label{tab:Models}
\end{table}

\newpage


\end{document}